\newif\ifdraft
\draftfalse
\newif\ifpreparepdf                       
\preparepdftrue 
\newif\ifhighlightedits
\highlighteditsfalse

        \ifdraft
\documentclass[aps,preprint,showpacs,longbibliography]{revtex4-1} 
        \else
\documentclass[aps,showpacs,superscriptaddress,longbibliography]{revtex4-1} 
        \fi

\ifpreparepdf
    \usepackage{color} 
    \usepackage[colorlinks]{hyperref} 
\else 
\fi

\usepackage{graphicx,amssymb,amsbsy}
\usepackage{ifthen}
\usepackage[percent]{overpic}
\usepackage{hyperref}
\usepackage{amsmath}

\graphicspath{{figs/}}  

\newcommand{\Sone}{\ensuremath{\textrm{S1}}}
\newcommand{\SoneN}{\ensuremath{\textrm{S1N}}}
\newcommand{\eval}[2]{\ensuremath{\lambda^{#1}_{#2}}} 
\newcommand{\evec}[2]{\ensuremath{V^{#1}_{#2}}} 
\newcommand{\evecRed}[2]{\ensuremath{\hat{V}^{#1}_{#2}}} 
\newcommand{\evecRRed}[2]{\ensuremath{\tilde{V}^{#1}_{#2}}} 

\ifdraft    
   
   \newcommand{\PC}[1]{$\footnotemark\footnotetext{PC: #1}$}
   
   \newcommand{\NBB}[1]{$\footnotemark\footnotetext{NBB: #1}$}
   
   \newcommand{\APW}[1]{$\footnotemark\footnotetext{APW: #1}$}
   
   \newcommand{\KYS}[1]{$\footnotemark\footnotetext{KYS: #1}$}

   \newcommand{\file}[1]{$\footnotemark\footnotetext{{\bf file} #1}$}
   \newcommand{\mycomment}[2]{\noindent \textbf{\underline{#1}}: \emph{#2}}
\else   

   \newcommand{\PC}[1]{}
   \newcommand{\JFG}[1]{}
   \newcommand{\NBB}[1]{}
   \newcommand{\APW}[1]{}

   \newcommand{\KYS}[1]{}
   
   \newcommand{\file}[1]{}
   \newcommand{\mycomment}[2]{}

\fi 

\ifhighlightedits
\else
\fi

\newcommand{\arXiv}[1]{\href{http://arXiv.org/abs/#1}{arXiv:#1}}

\newcommand{\rf}     [1] {~\cite{#1}}
\newcommand{\rfp}[1] {~\citep{#1}}
\newcommand{\refref} [1] {ref.~\cite{#1}}

\newcommand{\refrefs}[1] {refs.~\cite{#1}}

\newcommand{\refeq}  [1] {(\ref{#1})}
\newcommand{\refeqs} [2]{(\ref{#1}--\ref{#2})}

\newcommand{\reffig} [1] {figure~\ref{#1}}

\newcommand{\refFig} [1] {Figure~\ref{#1}}

\newcommand{\refsect}[1] {section~\ref{#1}}

\newcommand{\beq}{\begin{equation}}
\newcommand{\continue}{\nonumber \\ }

\newcommand{\eeq}{\end{equation}}
\newcommand{\ee}[1] {\label{#1} \end{equation}}

\newcommand{\bea}{\begin{eqnarray}}
\newcommand{\eea}{\end{eqnarray}}
\newcommand{\barr}{\begin{array}}
\newcommand{\earr}{\end{array}}

\newcommand{\etal}{{\em et al.}}    
\newcommand{\ie}{{i.e.}}        

\newcommand{\fFslice}{first Fourier mode slice}

\newcommand{\zeit}{\ensuremath{\tau}}  
\newcommand{\sliceTan}[1]{\ensuremath{t{}'_{#1}}}    
\newcommand{\groupTan}[1]{\ensuremath{t_{#1}}}    
\newcommand{\Lg}{\ensuremath{T}}   
\newcommand{\LieEl}{\ensuremath{g}}  
\newcommand{\LieElz}[1]{\ensuremath{\LieEl_z (#1)}}  
\newcommand{\LieEltheta}[1]{\ensuremath{\LieEl_\theta (#1)}}  
\newcommand{\id}{{\ \hbox{{\rm 1}\kern-.6em\hbox{\rm 1}}}}

\newcommand{\On}[1]{\ensuremath{\textrm{O}(#1)}}
\newcommand{\SOn}[1]{\ensuremath{\textrm{SO}(#1)}}         
\newcommand{\sspRed}{\ensuremath{\hat{\ssp}}}    
\newcommand{\sspRRed}{\ensuremath{\tilde{\ssp}}}    
\newcommand{\velRed}{\ensuremath{\hat{\vel}}}    
\newcommand{\velRRed}{\ensuremath{\tilde{\vel}}}    
\newcommand{\slicepz}{\ensuremath{\hat{\ssp}'}}   
\newcommand{\sliceptheta}{\ensuremath{\tilde{\ssp}'}}   
\newcommand{\Group}{\ensuremath{G}}         

\newcommand{\NSe}{Navier--Stokes equations}

\newcommand{\Reynolds}{\textit{Re}}  

\newcommand{\Statesp}{State space}
\newcommand{\statesp}{state space}

\newcommand\Real{\mbox{Re}\,} 
\newcommand\Imag{\mbox{Im}\,} 

\newcommand\flow[2]{{f^{#1}(#2)}}

\newcommand{\ssp}{\ensuremath{a}}            
\newcommand{\vel}{\ensuremath{v}}   

\newcommand{\eigExp}[1][]{
\ifthenelse{\equal{#1}{}}{\ensuremath{\lambda}}{\ensuremath{\lambda^{(#1)}}}
                        }
\newcommand{\eigRe}[1][]{
\ifthenelse{\equal{#1}{}}{\ensuremath{\mu}}{\ensuremath{\mu^{(#1)}}}
                        }
\newcommand{\eigIm}[1][]{
  \ifthenelse{\equal{#1}{}}{\ensuremath{\omega}}{\ensuremath{\omega^{(#1)}}}
            }

\newcommand{\inprod}[2]{\left\langle #1 ,\, #2 \right\rangle}

\begin{document}

\title{
Complexity of the laminar-turbulent boundary in pipe flow 
}
\author{Nazmi Burak Budanur}
\affiliation{Nonlinear Dynamics and Turbulence Group,
             IST Austria,
             3400 Klosterneuburg, Austria}
\affiliation{Kavli Institute for Theoretical Physics,
             UC Santa Barbara, Santa Barbara, CA 93106}

\author{Bj\"{o}rn Hof}
\affiliation{Nonlinear Dynamics and Turbulence Group,
             IST Austria,
             3400 Klosterneuburg, Austria}

\date{\today}

\begin{abstract}
Over the past decade, the edge of chaos has proven to be a 
fruitful starting point for investigations of shear flows
when the laminar base flow is linearly stable. 
Numerous computational studies of shear flows 
demonstrated the existence of states
that separate laminar and turbulent regions of the 
\statesp. 
In addition, some studies 
determined invariant solutions that reside on this edge. 
In this paper, we study the unstable manifold of one such solution
with the aid of continuous symmetry-reduction, which
	  we formulate here for the first time for the simultaneous 
	  quotiening of axial and azimuthal symmetries. 
Upon our investigation 
of the unstable manifold,
we discover a previously unknown traveling wave 
solution on the laminar-turbulent boundary with a relatively complex
structure.
By means of low-dimensional projections, we visualize different 
dynamical paths that connect these solutions to the turbulence. 
Our numerical experiments demonstrate that the laminar-turbulent 
boundary exhibits qualitatively different regions whose properties 
are influenced by the nearby invariant solutions. 
\end{abstract}


\maketitle

\ifdraft
\tableofcontents
\fi

\section{Introduction}
\label{s:intro}

Pipe flow is the most prominent member of a class of canonical
	shear flows where transition to turbulence occurs despite the 
	linear stability of the laminar state \rf{MesTre03}.
In the past two decades this problem enjoyed several major developments. 
The discovery of nonlinear traveling wave 
solutions \rf{FE03,Pringle07,Pringle09} and studies 
\rf{SchEckYor07,duguet07,MMSE09} of the laminar-turbulent boundary began 
to elucidate the \statesp\ of the system using insights from
dynamical systems theory.

In the dynamical systems approach to turbulence, fluid motion is 
envisioned as a trajectory in an infinite-dimensional 
\statesp \rfp{hopf48}. For the case of shear flows with linearly 
stable laminar solutions, this \statesp\ accommodates a stable
equilibrium point corresponding to the laminar solution and a 
chaotic set (an attractor or repellor) that is turbulence. 
Once this viewpoint is established, a natural question to ask is
what separates these two distinct regions in the \statesp .
In a small computational cell of channel flow,
Itano and Toh \rf{IT01} were first to study solutions
that neither laminarize nor become turbulent using 
a shooting method
and they discovered such trajectories tend towards a traveling 
wave solution. Following studies of shear flows in similar computational domains
found
periodic-like \rfp{TI03}, 
equilibrium \rfp{SGLDE08},
as well as seemingly chaotic \rfp{SchEckYor07} 
solutions using similar shooting methods. These findings
suggested the following picture of the 
\statesp\ of shear flows: The basin boundary between laminar 
and turbulent solutions is the stable manifold of an invariant set, 
whose unstable manifold on one side 
connects to the laminar solution 
and on the other side to the turbulent part of the \statesp . 
Schneider \etal\ 
\rf{SchEckYor07} named this invariant set the ``edge state''. 

Schneider \etal\ \rf{SchEckYor07} found in a short 
axially-periodic computational domain of pipe flow that 
the edge state exhibits chaotic 
motion, with flow structures much simpler than those of turbulence:
a slow streak in the center surrounded by two fast streaks with 
chaotically moving streamwise vortices in between.
A similar investigation in a long computational domain of pipe flow
\rf{MMSE09} yielded stream-wise localized edge states with chaotic
dynamics, whose flow 
fields at the core of the localized structure
resembled those computed in the short computational domain.

Typically when flows are not simplified by additional 
      symmetries the edge state tends to be chaotic as is the case for 
      plane Couette\rf{ScMaEc10}, channel\rf{ZamEck14a}, 
      and asymptotic suction boundary layer\rf{KKSDEH2016} flows.
In this seemingly generic case, the definition of the edge state is 
less clear than when it is formed by an exact invariant solution 
such as an (relative) equilibrium or a (relative) periodic orbit. 
In contrast to the invariant solutions, it is not straightforward to 
define and compute the stable/unstable manifolds of chaotic 
solutions, hence their theoretical study is much more challenging.

One way of systematically investigating chaotic edge states is 
studying the unstable invariant solutions that are contained within, 
since the geometry of chaotic sets are influenced by the invariant 
solutions that are embedded in them\rf{DasBuch}.
This strategy was adopted by Duguet \etal\ \rf{duguet07}, who found
for the case of pipe flow in an approximately 5-diameters long, 
axially periodic computational domain at $\Reynolds = 2875$ 
(\Reynolds\ based on bulk velocity and pipe diameter)
that the chaotic edge state evolves around the ``asymmetric'' 
traveling wave solution found by Pringle and Kerswell\rf{Pringle07}. 
In addition, they found a rotating traveling wave that had flow 
structures very similar to those of the asymmetric wave, but 
differently it rotated in the azimuthal direction in addition to 
drifting downstream.

\begin{figure}
	\centering
	\includegraphics[height=0.5\textwidth]{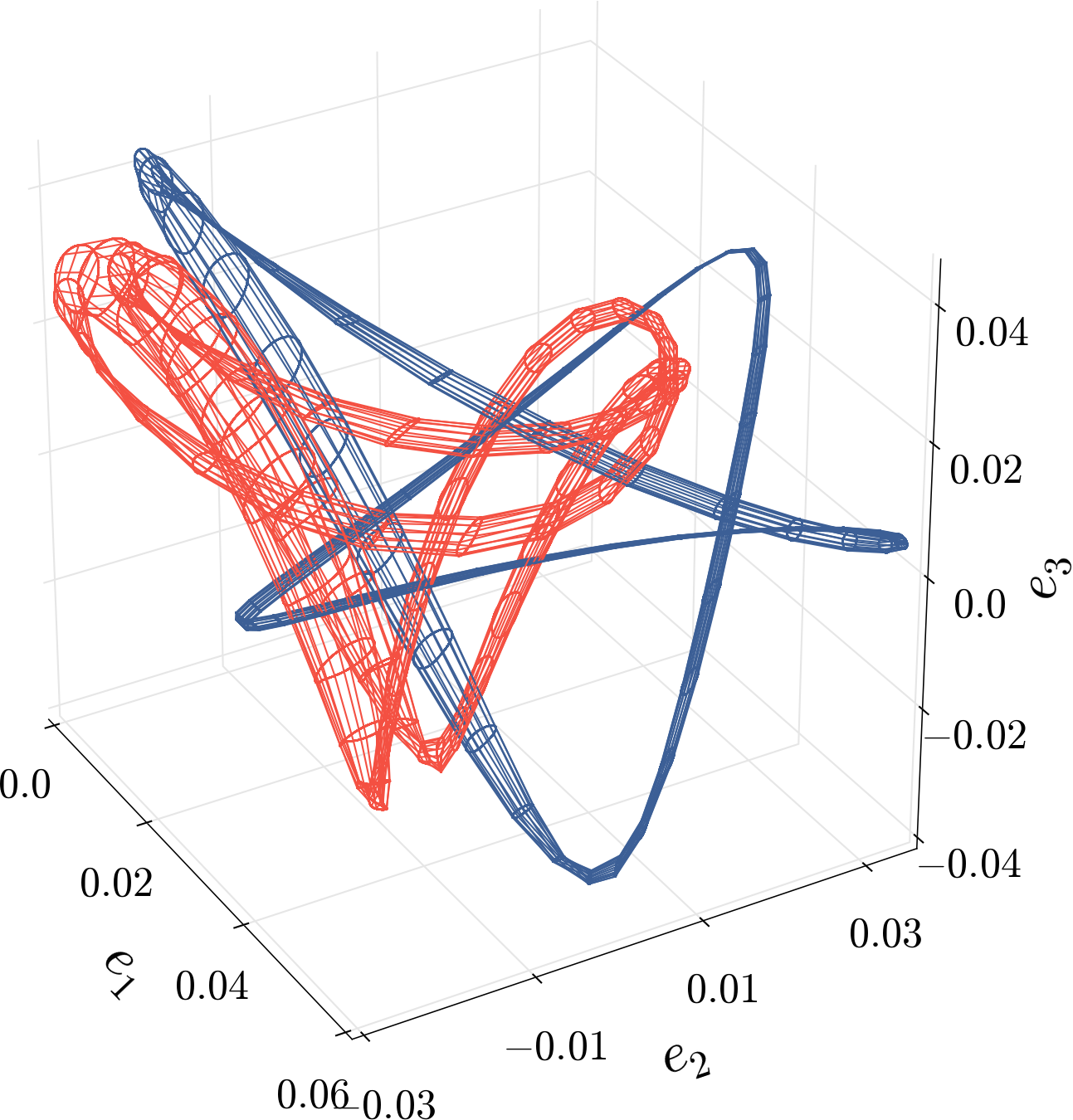} 
	\caption{
		Two traveling wave solutions of pipe flow
		(\Sone\ and \SoneN, to be described later in the article)
		and their continuum of symmetry copies obtained from 
		the axial translations and the azimuthal rotations, 
		visualized as red and blue 
		wire frames projected from the \statesp .
		Projection bases are the same with \reffig{f-man1winst}
		and \reffig{f-man3DS1}.		
		Apparent intersections of the curves are the artifacts of 
		finite-dimensional projection from the infinite-dimensional
		\statesp . In the symmetry-reduced \statesp , each torus will be 
		represented by a single point.
		\label{f-tori}
	}
\end{figure}

Canonical shear flows (pipe, plane Couette, and plane Poiseuille) 
are symmetric under continuous translations in
streamwise and spanwise (or azimuthal, in the case of pipe flow)
directions. This implies that every generic 
(non-symmetric) solution of these systems have infinitely many 
copies that can be generated by symmetry transformations. 
We 
visualized this degeneracy on \reffig{f-tori} for two traveling
wave solutions we study in this article.
Due to this multiplicity, something as simple as measuring the 
distance between 
two solutions becomes a daunting task. As a result, a common practice 
in dynamical systems approach to turbulence literature
is to use quantities averaged over computational domains, such as
energy input, dissipation, or pressure gradient, 
as indicators of ``closeness'' in the \statesp\ 
\rf{KawKida01,Visw07b,RiMeAv16,KKSDEH2016}. While observing such quantities can 
be used for deciding if two solutions are far from each other, they 
cannot be used to conclusively decide if two solutions are close in 
the \statesp : Two solutions that have similar rates of energy input,
dissipation, and pressure drop might have completely different flow 
structures. In this paper, this problem is addressed by symmetry
reduction, for the first time in both axial and azimuthal directions
of pipe flow. 
In the symmetry-reduced \statesp, 
	  the tori, such as those visualized on \reffig{f-tori}, 
	  are represented by single points, which 
	  simplifies the analysis substantially.
As we will explain later in the article, our 
approach to this problem is generic and adaptable to other 
canonical shear flow geometries in a straightforward fashion. 

For the application of our methods, we chose to revisit the 
laminar-turbulent boundary in a short ($\approx 5$ diameters) pipe
flow at $\Reynolds = 3000$. 
To this end, we visualize numerical approximations to the 
unstable manifold of the asymmetric traveling wave that reside in 
the laminar-turbulent boundary on local projections akin to those 
pioneered by Gibson \etal \rf{GHCW07}. 
We demonstrate that some parts of this unstable manifold that 
belong to the laminar-turbulent boundary exhibit dynamics 
qualitatively different from those previously attributed to the edge 
state in this setting.
Upon further investigation, we discover that this region is in the 
vicinity of a previously unknown traveling wave solution with four
high-speed streamwise streaks.  This new traveling wave appears to 
belong to a higher-energy region in the edge state; closer to the 
turbulent part of the \statesp. We study the unstable manifold of 
the new traveling wave and demonstrate its different connections 
to the turbulence. Finally, we present the results of numerical 
experiments, which demonstrate that the trajectories on 
laminar-turbulent boundary of pipe flow transiently approach to 
the traveling waves that reside in the edge.

The rest of the paper is organized as follows. In the next section, 
we overview our methods, particularly in \refsect{s:slices}, we 
present the generalization of the \fFslice \rfp{BudHof17} for
simultaneous reduction of continuous symmetries in axial
and azimuthal directions. We 
present our results in \refsect{s:results}, followed by the discussion
in \refsect{s:discus}.

\section{Methods}
\label{s:methods}
\subsection{Numerical set-up}

Numerical integration of the \NSe\
\beq
{\bf u}_\zeit
+ {\bf u}_{HP} \cdot \nabla {\bf u}
+ {\bf u} \cdot \nabla {\bf u}_{HP}
+ {\bf u} \cdot \nabla {\bf u}
= - \nabla p + 32 \frac{\beta}{\Reynolds} {\bf \hat{z}}
+ \frac{1}{\Reynolds} \nabla^2 {\bf u} \, , \label{e-NSeF}
\eeq
are performed using \texttt{Openpipeflow}\rfp{Willis2017}.
The velocity field
${\bf u}(z,r,\theta;\zeit)$ denotes the deviations from the
base (Hagen-Poiseuille) solution
${\bf u}_{HP} (z, r, \theta) =  2 (1 - (2r)^2) {\bf \hat{z}}$. 
Lengths and velocities are nondimensionalized by the pipe diameter $D$, 
and the mean axial speed $U$. 
Boundary conditions are no-slip and impermeable on pipe walls
${\bf u}(r = 1/2) = 0$; periodic in  axial and azimuthal
directions
${\bf u}(z, r, \theta) = 
 {\bf u}(z + kL, r, \theta + m 2 \pi)\,,\quad k, m \in \mathbb{Z}$. 
The velocity field satisfies the incompressibility condition 
$\nabla \cdot {\bf u} = 0$ and 
$\beta = \beta ({\bf u} (\zeit))$ is a feedback term, adjusted
in order to ensure a constant-flux equal to
that of the laminar solution at a given $\Reynolds$.
For all results of this paper,
$\Reynolds = U D / \nu = 3000$; the pipe length is set to 
$L= \pi / 0.625 \approx 5$; 
flow fields are discretized using $N = 128$ finite-difference points in 
radial direction and Fourier series truncated at 
$K = M = 64$ respectively in axial and azimuthal directions. 
Nonlinear terms are evaluated in the physical space on 
$(N \times 3K \times 3M)$ grid points following the $3/2$-rule for
dealiasing in Fourier-expanded directions. This truncation yields 
more than $6 \times 10^{6}$ numerical 
degrees of freedom, which is significantly higher than the typical 
resolutions used in similar computational studies 
\rf{SchEckYor07,WK04}. Our choice yields at least $5$ orders of 
magnitude drop in the spectral coefficients of the turbulent 
solutions at $\Reynolds = 3000$. While the solutions in the edge 
state generally require much less degrees of freedom to be
resolved, we made this ``conservative'' choice since we 
investigate different regions of the edge state and we did not know 
a priori the maximum resolution requirement.

\subsection{Symmetries}
	\label{s:symmPipe}

In this section, we review the symmetries of pipe flow and 
   	  their representations in terms of their actions on the 
   	  velocity field. For more detailed discussions of the 
   	  symmetries and the invariant subspaces of pipe flow, 
   	  we refer the reader to \refrefs{WFSBC15,ACHKW11}.
Pipe flow is equivariant under 
the axial translations $\LieElz{l}, \, l \in [0, L)$, 
the azimuthal rotations $\LieEltheta{\phi}\, \phi \in [0, 2\pi)$, and 
the reflection $\sigma$; whose actions on the 
axial $u$, radial $v$, and azimuthal $w$ components of the
velocity field are 
given by
\bea
	\LieElz{l} [u, v, w] (z, r , \theta) &=& 
		[u, v, w] (z - l, r, \theta) \, , \label{e-translation} \\ 
	\LieEltheta{ \phi } [u, v, w] (z, r , \theta) &=& 
	[u, v, w] (z, r, \theta - \phi) \, ,  \label{e-rotation} \\
	\sigma [u, v, w] (z, r , \theta) &=& 
	[u, v, - w] (z, r, - \theta) \, .  \label{e-reflection}
\eea
Therefore, the symmetry group of pipe flow is the direct product of
$\SOn{2}_z$ and $\On{2}_\theta$, \ie\ 
\beq 
\Group = \SOn{2}_z \times \On{2}_\theta = 
\{\LieElz{l}, \LieEltheta{ \phi }, \sigma\}. \label{e-Group}
\eeq

Lie group actions \refeq{e-translation} and \refeq{e-rotation} can
be written as operator-exponentials of their respective 
infinitesimal generators 
$\Lg_z$ and $\Lg_\theta$ as 
\beq
	\LieElz{l} = e^{\Lg_z l} \quad \mbox{and} \quad 
	\LieEltheta{\phi} = e^{\Lg_\theta \phi} \, ,
\eeq
where the actions of $\Lg_z$ and $\Lg_\theta$ on the velocity field 
${\bf u} = {\bf u} (z, r, \theta)$
are 
\beq
	\Lg_z {\bf u} = - \frac{\partial}{\partial z} {\bf u} 
	\quad \mbox{and} \quad 
	\Lg_\theta {\bf u} = - \frac{\partial}{\partial \theta} {\bf u} 
	\, .
	\label{e-Lg}
\eeq

\subsection{\Statesp\ notation}
\label{s:statesp}

Since we are going to use dynamical systems tools, it is handy to 
introduce a \statesp\ notation for use in the rest of the paper. 
Let $\ssp (0)$ 
be a vector that contains all numerical degrees of freedom of a 
three-dimensional velocity field ${\bf u} (z, r, \theta; 0)$ at an initial time
$\zeit = 0$. Then, the \NSe\ 
\refeq{e-NSeF} along with the incompressibility and the 
boundary conditions 
imply a finite-time flow
\beq
	\ssp (\zeit) = \flow{\zeit}{\ssp (0)} \, , \label{e-flow}
\eeq
where $\ssp (\zeit)$ corresponds to the velocity field 
${\bf u} (z, r, \theta; \zeit)$ at time $\zeit$. Assuming that the flow 
\refeq{e-flow} is smooth, we can also represent the system
as a high-dimensional ordinary differential equation
\beq
	\dot{\ssp} = \vel(\ssp) 
			   = \lim_{\delta \zeit \rightarrow 0} 
				  (\flow{\delta \zeit}{\ssp} - \ssp)
				 / \delta \zeit \, .
\eeq

Actions of group elements $\LieEl \in \Group$ on a \statesp\ 
vector $\ssp$ should be thought as actions on the corresponding 
velocity fields as in \refeqs{e-translation}{e-reflection}. 
In other words, 
if $\ssp$ corresponds to velocity field ${\bf u}$ 
then $\LieEl \ssp$ corresponds to the transformed velocity field
$\LieEl {\bf u}$. Similarly, group tangents 
$\groupTan{z, \theta} (\ssp) = \Lg_{z, \theta} \ssp$ correspond
to velocity fields $\Lg_{z, \theta} {\bf u}$, where $\Lg_{z, \theta}$ 
acts as in \refeq{e-Lg}. In the \statesp ,
equivariance under $\LieEl$ implies that \statesp\ velocity and
finite-time flow commutes with $\LieEl$, \ie\ 
\beq
	\vel(\LieEl \ssp) = \LieEl \vel(\ssp)\,,\quad \mbox{and} \,,\quad 
	\flow{\zeit}{\LieEl \ssp} = \LieEl \flow{\zeit}{\ssp}.
	\label{e-Equivariance}
\eeq

Let $\ssp$ and $\ssp'$ correspond to velocity fields 
${\bf u}$ and ${\bf u}'$ respectively, we define the inner product
\beq
	\inprod{\ssp}{\ssp'} = 
	\frac{1}{2} \int {\bf u} \cdot {\bf u'} \, dV \, , 
	\label{e-Norm}
\eeq
where the integral is carried over the pipe volume. Thus, 
$||\ssp||^2 = \inprod{\ssp}{\ssp}$ gives the kinetic energy
of the velocity fluctuations; hence this choice of norm is usually
referred to as the ``energy norm''.

\subsection{Continuous symmetry reduction}
\label{s:slices}

We define the group orbit $\mathcal{M}_\ssp $ of a 
\statesp\ point $\ssp$ as all 
points reachable from $\ssp$ by symmetry transformations, \ie\ 
\beq
	\mathcal{M}_\ssp = 
	\left\{ \LieEl \ssp\, |\, \LieEl \in \Group \right\} \, .
	\label{e-gOrbit}
\eeq
If $\ssp$ is not invariant under any symmetries of the system, its 
group orbit \refeq{e-gOrbit} defines two distinct two-tori that are
related to each other by the reflection $\sigma$ since we have two 
compact continuous symmetry directions \refeq{e-Group}. All \statesp\
points on a group orbit have the same physical properties, such as 
kinetic energy, dissipation, or wall-friction since these quantities
are invariants of symmetry transformations. In addition, the dynamics
of each point on a group orbit can be obtained from the dynamics of 
a single point following the definition \refeq{e-Equivariance} of
equivariance. In other words, the \statesp\ of pipe flow exhibits a lot
of redundancy since a generic point have infinitely many symmetry 
copies. Moreover, the presence of continuous symmetries renders the
study of \statesp\ extremely hard: 
in the presence of pipe flow's symmetries, 
measuring the distance between two generic \statesp\ points 
becomes the question of the minimum distance between tori, 
whose computational cost can easily
become prohibitive if it is 
to be carried out repeatedly.
\textit{Continuous symmetry reduction}, which we introduce next,  
is a coordinate transformation such that 
\statesp\ points that are related by a continuous symmetries are 
represented by a single point in the reduced \statesp. 

We begin by reducing the streamwise translation symmetry following
\rf{BudHof17} exactly: We define a
``slice template'' $\slicepz$ with a corresponding 
three-dimensional velocity field, 
whose each component 
$\hat{{\bf u}}'_k$ is defined as
\beq
\hat{{\bf u}}'_k(z, \theta , r) 
= J_0(\alpha r) \cos (2 \pi z / L) \,,\quad\, k = 1, 2, 3,
\label{e-slicepz}
\eeq
where $J_0$ is the Bessel function of the first kind, which vanishes
at the pipe wall, \ie\ $J_0 (\alpha / 2) = 0$. Then the translation
symmetry-reduced coordinates are given by
\bea
	\sspRed (\zeit) &=& \LieEl_z (L \phi_z / 2 \pi  ) \ssp (\zeit) 
	\,, \, \mbox{where} \label{e-sspred} \\
	\phi_z (\zeit) &=& \arg (\inprod{\ssp(\zeit)}{\slicepz} + i 
	\inprod{\ssp(\zeit)}{\LieEl_z(- L/4) \slicepz})
	\,. \label{e-slicePhasez} 
\eea
The transformation \refeq{e-sspred} exists as long as the phase 
\refeq{e-slicePhasez} does. Note that $\phi_z$ is the polar angle
when the state $\ssp (\zeit)$ is projected onto the plane spanned 
by $(\slicepz, \LieEl_z(- L/4) \slicepz)$. 

Extension of \refeq{e-sspred} 
for the azimuthal symmetry reduction is straightforward with a second
slice template \sliceptheta\ corresponding to the velocity field 
with components 
\beq
\tilde{{\bf u}}'_k(z, \theta , r) 
= J_0(\alpha r) \cos (\theta) \,,\quad\, k = 1, 2, 3\,.
\label{e-sliceptheta}
\eeq
Then the symmetry-reducing coordinate transformation becomes
\bea
\sspRRed (\zeit) &=& \LieEl_\theta (\phi_\theta) \ssp (\zeit) 
\,, \, \mbox{where} \label{e-ssprred} \\
\phi_\theta (\zeit) &=& 
	     \arg (\inprod{\sspRed(\zeit)}{\sliceptheta} 
       + i \inprod{\sspRed(\zeit)}{
			       \LieEl_\theta(- \pi/2) \sliceptheta})
\,. \label{e-slicePhasetheta} 
\eea
Similar to \refeq{e-sspred} and \refeq{e-slicePhasez}, 
the transformation \refeq{e-ssprred} exists as long as the phase
\refeq{e-slicePhasetheta} does. Our choice of the order at which
the continuous symmetries are reduced is merely a convention 
since the inner products in 
\refeq{e-slicePhasez} and \refeq{e-slicePhasetheta} 
remain unchanged respectively under transformations
\refeq{e-ssprred} and \refeq{e-sspred}. 
This is a result of 
our particular choice of the slice templates 
	 \refeq{e-slicepz} and \refeq{e-sliceptheta}, which respectively 
	 do not depend on $\theta$ and $z$ and
the fact
that the axial translation and the azimuthal rotation 
symmetries commute.
For a general symmetry group with non-commuting elements, a more 
careful treatment would have been required.

    The slice templates \refeq{e-slicepz} and 
	  \refeq{e-sliceptheta} need not be valid (smooth, divergence free) 
	  pipe flow velocity fields. 
  	  The only requirement on the slice templates $\slicepz$ and
  	  $\sliceptheta$
      is that the projections of the 2-torus
      $\{\LieEl_\theta (\phi) \LieEl_z(l) \ssp\,|\, l \in [0, L)\,, \phi \in [0, 2 \pi)\}$
      onto the $(\slicepz, \LieEl_z(- L/4) \slicepz)$-
  	  and 
  	  $(\sliceptheta, \LieEl_\theta(- \pi/2) \sliceptheta)$-planes 
      both must be circles for a generic state $\ssp$. Cosine-dependence
  	  of \refeq{e-slicepz} and \refeq{e-sliceptheta}  
  	  on the respective symmetry directions provides this as long as 
  	  the projection is nonzero.
	  We determine the rest of the slice templates by experimentation
	  in order to reduce the probability of having a vanishing 
	  projection:
  	  We decided to use all three velocity components for the 
  	  template-fields in order to receive contributions from all 
  	  directions.   	  
  	  The radial-dependence on $J_0(\alpha r)$, on the other hand, is 
  	  an arbitrary and it is conceivable that there could be other
  	  equally valid choices.
  	  As we will further argue after introducing the slice phase 
  	  velocities \refeq{e-phaseVels},       
  	  our experience with the slice templates
  	  \refeq{e-slicepz} and \refeq{e-sliceptheta} has been that
      the symmetry-reduction procedure that we described yields no 
      discontinuities when applied to the generic turbulent trajectories.

Budanur \etal \rf{BudCvi14} showed for a one-dimensional PDE with \SOn{2}
symmetry that polar coordinate transformations similar 
to \refeq{e-sspred} and \refeq{e-ssprred} 
can be reformulated as a ``slice hyperplane''.
A slice hyperplane is a set of points perpendicular to 
the group orbit of a slice
template and is transversally intersected by the group orbits 
of the \statesp\ points in a closed neighborhood of the 
template. 
While a general slice has a finite region of 
applicability, if the template is chosen such that its dependence
on the symmetry coordinate depends only on the first Fourier
mode, then the slice works for a semi-infinite domain 
(half-hyperplane) that covers all \statesp\ of interest, with 
a regularizable singularity in time. This method is named 
``\fFslice '' in \rfp{BudCvi14} and we refer 
the reader to \rfp{BuBoCvSi14} for a pedagogical introduction
to it.

Reduced coordinates
\refeq{e-sspred} satisfy the half-hyperplane equation
\beq
	\inprod{\sspRed - \slicepz}{\sliceTan{z}} \,, \quad
	\inprod{\groupTan{z} (\sspRed)}{\sliceTan{z}} > 0 \,,
	\label{e-SliceHplanez}
\eeq
where $\groupTan{z} (\ssp) = \Lg_z \ssp$ is the 
group tangent of a \statesp\ point \ssp\ and
$\sliceTan{z} = \groupTan{z} (\slicepz)$. Similarly, we 
can express the consequent transformation \refeq{e-ssprred} as 
another half-hyperplane in the streamwise symmetry-reduced 
\statesp\ as
\beq
	\inprod{\sspRRed - \sliceptheta}{\sliceTan{\theta}} \,, \quad
	\inprod{\groupTan{\theta} (\sspRRed)}{\sliceTan{\theta}} > 0 \,,
	\label{e-SliceHplanetheta}
\eeq
where similarly, 
$\groupTan{\theta} (\sspRed) = \Lg_\theta (\sspRed)$ and
$\sliceTan{\theta} = \groupTan{\theta} (\sliceptheta)$. 

The main advantage of the reformulations 
\refeq{e-SliceHplanez} and \refeq{e-SliceHplanetheta} is that from 
this perspective one is able to derive projection operators for 
transforming the tangent space of $\ssp$ to the slice. Let 
$\delta \ssp$ be a small perturbation to $\ssp$ in the full \statesp ,
and $\phi_z$ be the slice phase that brings $\ssp$ to slice as 
$\sspRed = \LieElz{\phi_z L / 2 \pi} \ssp$, then the small 
perturbation can be brought to the slice by the projection
\beq
	\delta \sspRed = \left (1 
						- \frac{\groupTan{z}(\sspRed) 
								\otimes \sliceTan{z}}{ 
								\inprod{\groupTan{z} (\sspRed)}{\sliceTan{z}}}
					\right ) \LieElz{\phi_z L / 2 \pi} \delta \ssp
	\label{e-Projz}
\eeq
Similarly, $\delta \sspRed$ is transformed to the second slice 
\refeq{e-SliceHplanetheta} as
\beq
	\delta \sspRRed = \left (1 
	- \frac{\groupTan{\theta}(\sspRRed) 
		\otimes \sliceTan{\theta}}{ 
		\inprod{\groupTan{\theta} (\sspRRed)}{\sliceTan{\theta}}}
	\right ) \LieEltheta{\phi_\theta} \delta \sspRed	\,, 
	\label{e-Projtheta}
\eeq
where $\phi_\theta$ is the slice-fixing phase 
\refeq{e-slicePhasetheta}, \ie\ 
$\sspRRed = \LieEltheta{\phi_\theta} \sspRed$. For the derivation of 
these projection operators, we refer to the appendix of 
\rfp{BudHof17}.

\begin{figure}
	\centering
	\includegraphics[height=0.5\textwidth]{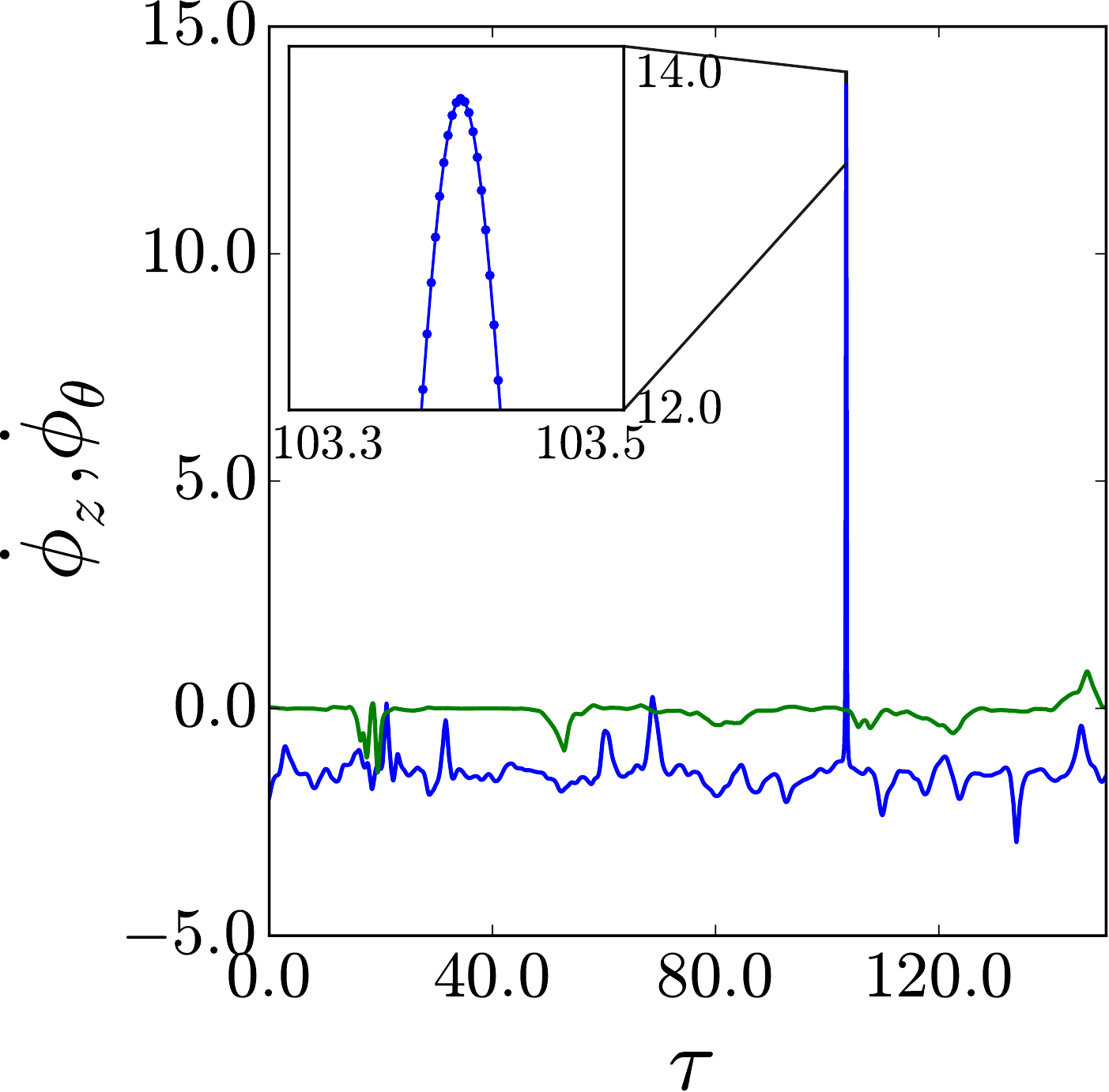} 
	\caption{Phase velocities
		\refeq{e-phaseVels}
		$\dot{\phi}_z$ (blue) and $\dot{\phi}_{\theta}$ (green)
		measured in a typical turbulence simulation. Inset:
		Zoom into the time interval
		$\zeit \in [103.3, 103.5]$, where a sharp increase in 
		$\dot{\phi}_z$ is apparent. Dots in the inset correspond 
		to adjacent time-steps of the simulation, showing that 
		this episode is in fact time-resolved.
		\label{f-dphidt}
	}
\end{figure}

We are going to finish this section by explaining possible short-comings
of the presented symmetry-reduction scheme: With the aid of the 
projections \refeq{e-Projz} and \refeq{e-Projtheta}, it is 
straightforward to express \statesp\ velocity in the reduced 
\statesp. Substituting $\delta \ssp$ with $\vel (\ssp)$
in \refeq{e-Projz} we obtain
\bea
	\velRed(\sspRed) &=&  
	\left (1 
	- \frac{\groupTan{z}(\sspRed) 
		\otimes \sliceTan{z}}{ 
		\inprod{\groupTan{z} (\sspRed)}{\sliceTan{z}}}
	\right ) \LieElz{\phi_z L / 2 \pi} \vel (\ssp) \, ,\continue
	\velRed(\sspRed) &=& 
	\vel (\sspRed) - \frac{ 
		\inprod{\sliceTan{z}}{\vel (\sspRed)}}{ 
		\inprod{\groupTan{z}(\sspRed)}{\sliceTan{z}}} 
		\groupTan{z}(\sspRed) \, ,
	\label{e-velRed}
\eea
where we used the first equivariance property in \refeq{e-Equivariance}
and defined the stream-wise symmetry-reduced \statesp\ velocity 
as $\velRed (\sspRed) = \dot{\sspRed}$. The fully 
symmetry-reduced 
\statesp\ velocity $\velRRed (\sspRRed) = \dot{\sspRRed}$
can be similarly written as
\beq
	\velRRed(\sspRRed) = 
	\velRed (\sspRRed) - \frac{ 
		\inprod{\sliceTan{\theta}}{\velRed (\sspRRed)}}{ 
		\inprod{\groupTan{\theta}(\sspRRed)}{\sliceTan{\theta}}} \groupTan{\theta}(\sspRRed) \, .
	\label{e-velRRed}	
\eeq
It is instructive to have a close look at 
\refeq{e-velRed}: The reduced \statesp\ velocity $\velRed$
is generated from full \statesp\ velocity $\vel$ by subtracting
the component in the direction of the group tangent with 
a pre-factor 
proportional to $\vel$'s projection onto the slice tangent. 
The second thing to recognize in \refeq{e-velRed} is the fact 
that $\inprod{\groupTan{z}(\sspRed)}{\sliceTan{z}}$ appears in the 
denominator, hence if this inner product vanishes, the reduced 
\statesp\ velocity field diverges. For a general slice, this 
condition sets the border in which the slice may apply 
\rf{FrCv11,atlas12}; 
in our particular case of the \fFslice ,
this condition is equivalent to the existence of the phase
\refeq{e-slicePhasez}. It can be shown\rf{rowley_reconstruction_2000} 
that the multipliers of group tangents in \refeq{e-velRed} and 
\refeq{e-velRRed} correspond to the time derivatives of the parameters
of group actions that bring trajectories to the slice, \ie\ 
\beq
	\dot{\phi}_z = \left(\frac{2 \pi}{L} \right) \frac{ 
		\inprod{\sliceTan{z}}{\vel (\sspRed)}}{ 
		\inprod{\groupTan{z}(\sspRed)}{\sliceTan{z}}} \,, \quad 
	\dot{\phi}_{\theta} = \frac{ 
		\inprod{\sliceTan{\theta}}{\velRed (\sspRRed)}}{ 
		\inprod{\groupTan{\theta}(\sspRRed)}{\sliceTan{\theta}}} \, .
	\label{e-phaseVels}
\eeq
Hence, these are appropriate quantities for checking whether or not the
sliced dynamics is within its borders. Our experience with the 
\fFslice s has been such that these quantities do not diverge for 
generic trajectories, although they may have ``fast'' episodes. We 
illustrate this for a typical turbulent trajectory in 
\reffig{f-dphidt}. Fast fluctuations that are caused by such episodes 
can be handled by rescaling the time-step of the numerical 
simulation or the problem can be explicitly reformulated using a 
``slice time'' \rf{BudCvi14}. In this study, we found a fixed 
time-step $\Delta \zeit= 0.0025$ to be sufficient to resolve the 
changes in slice phases, without resorting to an adaptive 
time-stepping scheme.

\section{Traveling waves on the laminar-turbulent boundary}
\label{s:results}
Other than the laminar equilibrium, all invariant solutions such as 
traveling waves,
rotating waves, relative periodic orbits, and possibly higher-dimensional 
invariant tori, of pipe flow 
drift downstream. 
A traveling wave $\ssp_{TW}$ is the simplest among those, 
which satisfies
\beq
    \ssp_{TW} (\zeit) = \LieElz{c \zeit} \ssp_{TW} (0) \, , 
\eeq
namely, its sole dynamics is a drift in the axial direction with 
constant phase speed $c$. Linear stability of this solution is 
described by the following eigenvalue problem \rf{ChossLaut00,DasBuch}
\beq
    \left( \left.\frac{d \vel (\ssp)}{d \ssp}\right|_{\ssp = \ssp_{TW}} 
         - c \Lg_z \right) \evec{TW}{i} = \eval{TW}{i} \evec{TW}{i}  \, ,
    \label{e-TWstability}
\eeq
where \eval{TW}{i} and \evec{TW}{i} are the linear
stability eigenvalues and eigenvectors
respectively. Since the \statesp\ is formally infinite-dimensional, each 
traveling wave has infinitely many stability eigenvalues and eigenvectors. In
practice, we solve \refeq{e-TWstability} by Arnoldi iteration and approximate 
the finite-dimensional leading (most unstable) part of the tangent space. 

Traveling waves become equilibria when the axial translation symmetry 
is reduced, say by the \fFslice\ method of \refeq{e-sspred}. The
translation symmetry-reduced \statesp\ still exhibits the 
$\On{2}_{\theta}$ and it follows from the normal form analysis that
all equilibria of an \On{2}-equivariant system belong to invariant 
subspaces of reflection symmetry or its conjugates\rf{AGHks88}. Coming 
back to full \statesp , this implies that all traveling waves of pipe 
flow must be invariant under the reflection $\sigma$ or a related 
symmetry.
In particular, the solutions we are going to 
investigate in what follows are invariant under so-called shift-and-reflect 
symmetry
\beq
    S = \sigma \LieEl_z (L / 2) \in \Group \, . \label{e-shiftnref}
\eeq
In the shift-and-reflect subspace, continuous rotation symmetry \refeq{e-rotation}
of the pipe flow is broken and only the discrete rotation by $\pi$ is allowed
\rfp{WFSBC15}. Thus, the symmetry group of shift-and-reflect subspace is 
\beq
    \Group_S = \{ \LieElz{l} \,, \LieEltheta{\pi} \} \,.
\eeq
Note that by definition, reflection symmetry is equivalent to an 
axial-translation by $L/2$. 

\subsection{Unstable manifold of \Sone}

\begin{figure}[h]
	\centering
	(a) \includegraphics[height=0.4\textwidth]{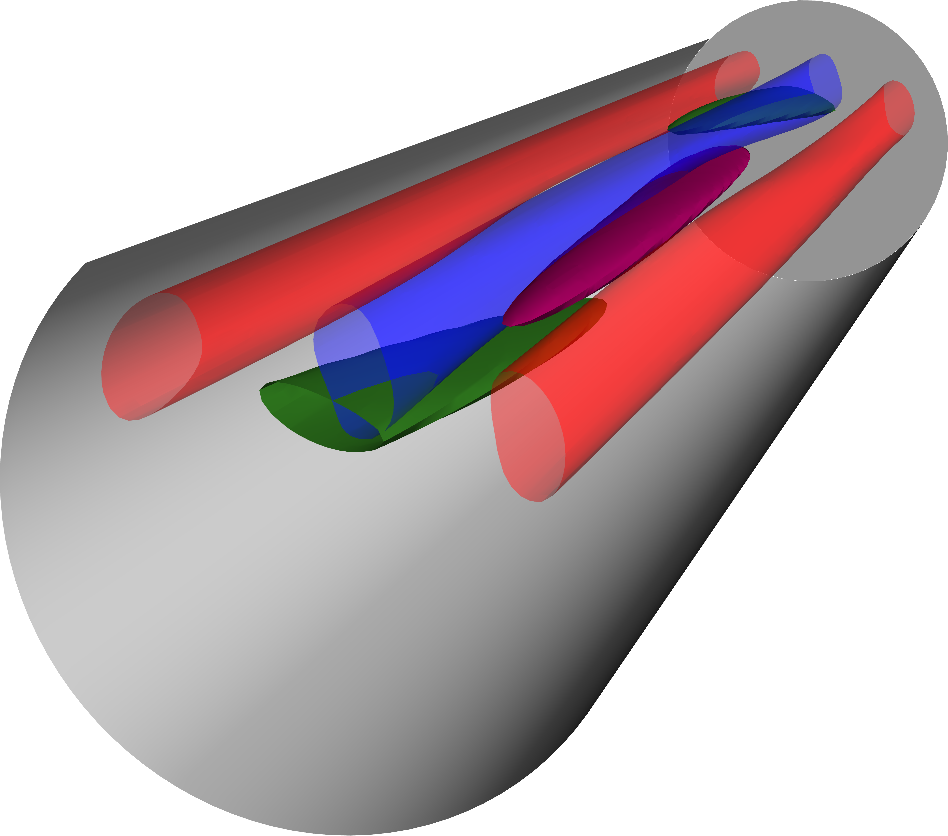} 
	(b) \includegraphics[height=0.4\textwidth]{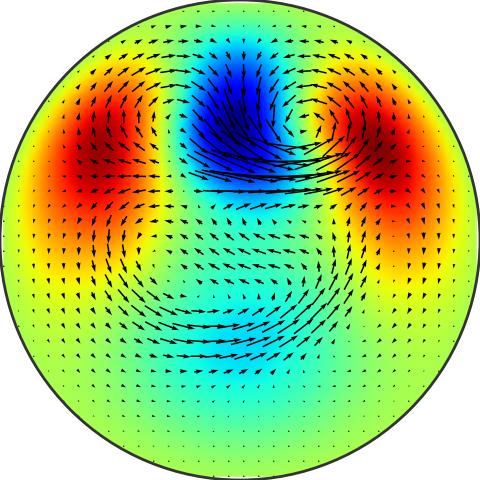} \\
	(c) \includegraphics[width=0.2\textwidth]{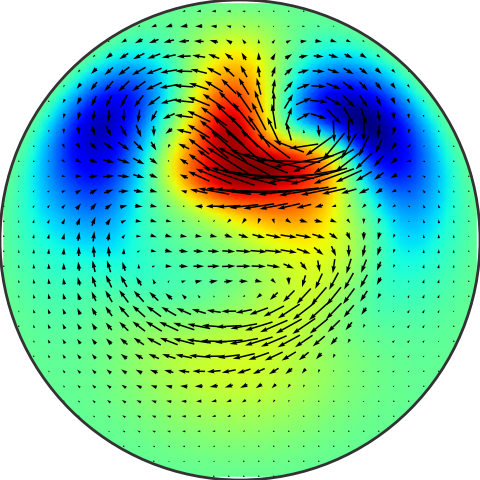} 
	(d) \includegraphics[width=0.2\textwidth]{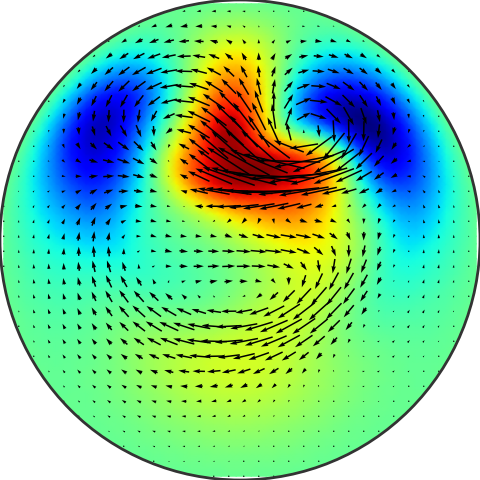} 
	(e) \includegraphics[width=0.2\textwidth]{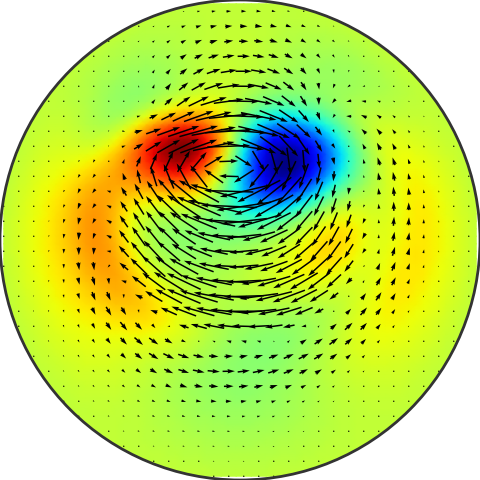} 
	(f) \includegraphics[width=0.2\textwidth]{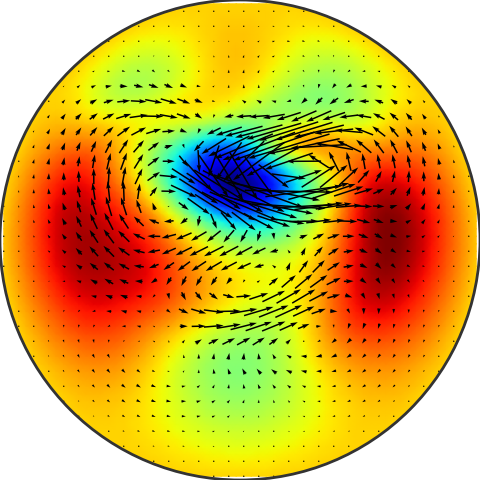} 
	\caption[\Sone\ and its Floquet vectors]{
    (a) Isosurfaces of streamwise velocity at 
    $u = 0.26$ (red) and $u = -0.33$ (blue) and
    streamwise vorticity at $\omega = \pm 0.40$ (green/purple)
    for the travelling wave \Sone\ in three dimensions. 
    Color-coded streamwise velocity and 
    cross-stream velocity (arrows) averaged over 
    half pipe-length $z \in (0, L/2]$ for
    \Sone\ (b), 
    and the unstable eigenvectors
    $\evec{\Sone}{1}$ (c), 
    $\evecRed{\Sone}{1}$ (d), 
    $\evec{\Sone}{2}$ (e), 
    $\evecRed{\Sone}{2}$ (f). Flow direction into the page.
    \label{f-S1}
	}
\end{figure}

Numerical\rfp{duguet07} and experimental\rfp{deLMeAvHo12} 
(combined with numerical work) 
evidence strongly suggests that the edge 
state of axially periodic pipe flow that is not long enough to exhibit 
streamwise localization 
is organized around the asymmetric traveling wave solution found in
\rf{Pringle07}. Following \rf{Pringle09}, we 
are going to refer to this solution as $\Sone$. This naming refers to 
solution's symmetries: $S$ stands for `shift-and-reflect' \refeq{e-shiftnref} 
invariance and $1$ is the fundamental azimuthal wave number. 
$S1$ appears at a low-\Reynolds\ through a symmetry-breaking 
bifurcation of a $\LieEltheta{\pi} \LieElz{L/2}$-symmetric solution, 
which itself is born out of a
saddle-node bifurcation at even lower \Reynolds \rf{Pringle07}. For the parameters studied
here ($\Reynolds = 3000, L = \pi / 0.625 \approx 5$) \Sone\ and its two 
purely real unstable eigenvectors 
with corresponding eigenvalues 
$\eval{\Sone}{1}  = 0.0793, 
\eval{\Sone}{2} = 0.0223$ are 
visualized in \reffig{f-S1}. 
On \reffig{f-S1} (a) and throughout the three-dimensional 
visualizations of this paper,
streamwise velocity isosurfaces
are chosen as $75\%$ of their maximum/minimum
values for velocity fluctuations. Similarly, vorticity isosurface 
levels are chosen at $60\%$ of their respective maxima and minima. 
Numerical values of velocity/vorticity isosurfaces are given in the 
figure caption. For the two-dimensional cross-sectional visualizations of 
shift-and-reflect-symmetric states, we chose to average quantities
over the first half pipe-length 
since the second half is simply the reflection
of the first. 

    In \reffig{f-S1},
	besides the eigenvectors \evec{\Sone}{1,2} (\reffig{f-S1} c and e),
	which are computed in the full \statesp ,
	we also show \evecRed{\Sone}{1,2} (\reffig{f-S1} d and f), which are 
	obtained by the projection \refeq{e-Projz}. With these visualizations, 
	we would like to emphasize that the
	symmetry-reduced eigenvector might be quite different from the one that is 
	computed in the full \statesp\ as it is the case for 
	\evec{\Sone}{2} (\reffig{f-S1} e) and \evecRed{\Sone}{2}(\reffig{f-S1} f).
	Moreover, it should be understood that the
	unstable manifold of \Sone\ in the full \statesp\ is three-dimensional 
	whose linear part is contained within the tangent space spanned by
	$\evec{\Sone}{1}$, $\evec{\Sone}{2}$,  
	and the marginal direction
	$\evec{\Sone}{3} = \groupTan{z}(\ssp_{\Sone}) = \Lg_z \ssp_{\Sone}$ 
	with the eigenvalue $\eval{\Sone}{3} = 0$. 
	Note that the projection \refeq{e-Projz} subtracts components in 
	this direction and its action on $\groupTan{z}(\ssp_{\Sone})$ 
	yields a zero vector. 
	In other words, the marginal stability direction, which corresponds 
	to axial drift of the traveling wave, is eliminated by slicing.
	Thus  within the slice, \Sone\ has a two-dimensional unstable manifold, 
	whose linear part is contained in the 
	$(\evecRed{\Sone}{1}, \evecRed{\Sone}{2})$-plane.

Duguet \etal \rf{duguet07} studied the evolution of perturbations along the 
unstable directions
of \Sone. After confirming that the perturbations in $\pm \evec{\Sone}{1}$ 
direction either laminarizes or develops into turbulence, they focused 
on the perturbations on $(\evec{\Sone}{1}, \evec{\Sone}{2})$ 
plane that neither laminarize nor 
become turbulent. Running Newton searches near-recurrences of these 
trajectories, they found a rotated versions of \Sone\ and conjectured 
that \Sone 's unstable manifold contains a ``relative''-heteroclinic connection 
to its rotation by
approximately $52^o$. This cannot be true since the unstable eigenvectors 
$\evec{\Sone}{1}$ and $\evec{\Sone}{2}$ 
are also shift-and-reflect symmetric, hence the associated 
unstable manifold lies in the shift-and-reflect invariant subspace, which 
only allows for azimuthal rotations by $\pi$. We are guessing that 
Duguet \etal's conjecture was a result of a numerical error build up due to not
restricting dynamics into the shift-and-reflect invariant subspace.
Our first calculation in this paper will be very similar to theirs, 
with the symmetry-restriction requirement taken into account. In addition, we 
are going to carry out our computation in the translation symmetry-reduced 
\statesp\ \refeq{e-sspred}, which will allow us to visualize the unstable 
manifold. 

\begin{figure}[h]
	\begin{minipage}{0.64\textwidth}
        (a) \includegraphics[height=0.9\textwidth]{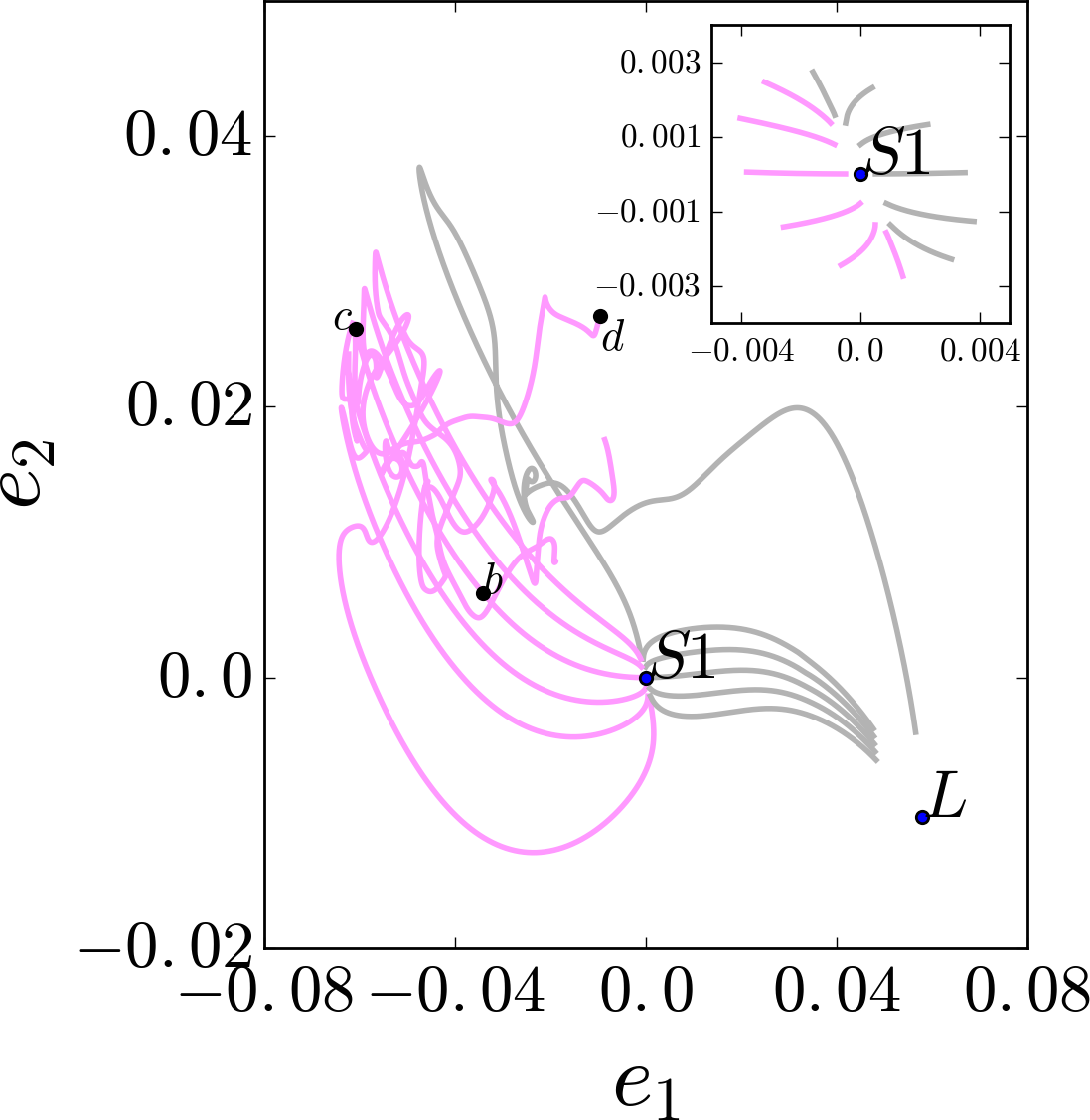} 
	\end{minipage}
	\begin{minipage}{0.31\textwidth}
		\begin{flushright}
		(b) \includegraphics[height=0.6\textwidth]{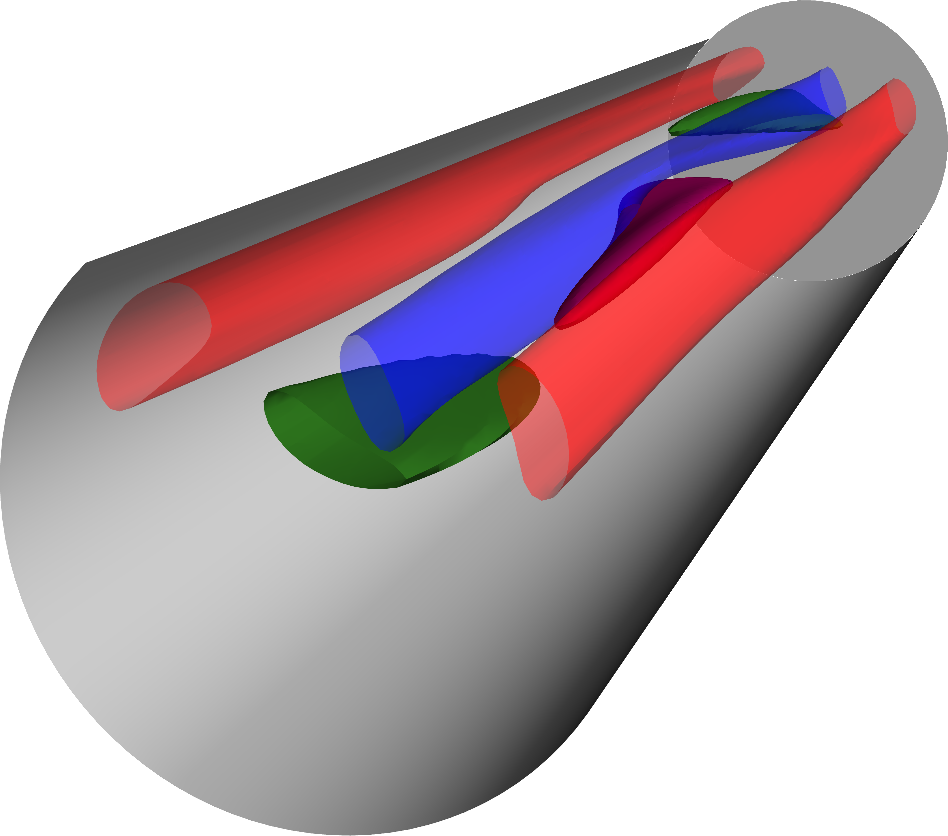} \\
		(c) \includegraphics[height=0.6\textwidth]{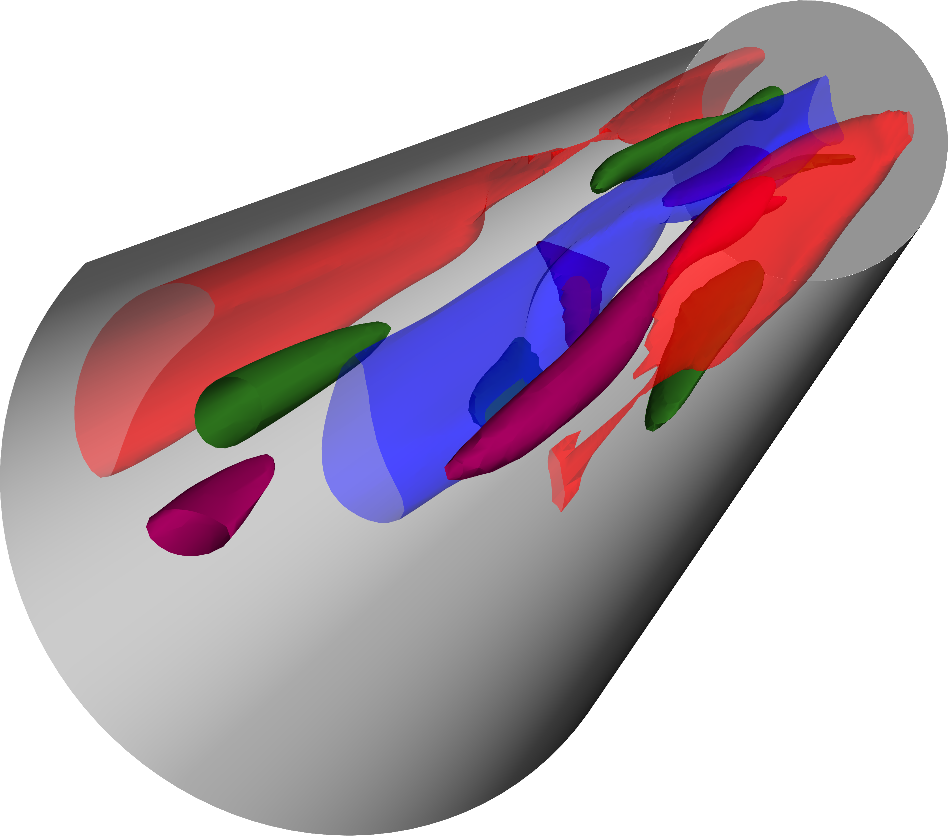} \\
		(d) \includegraphics[height=0.6\textwidth]{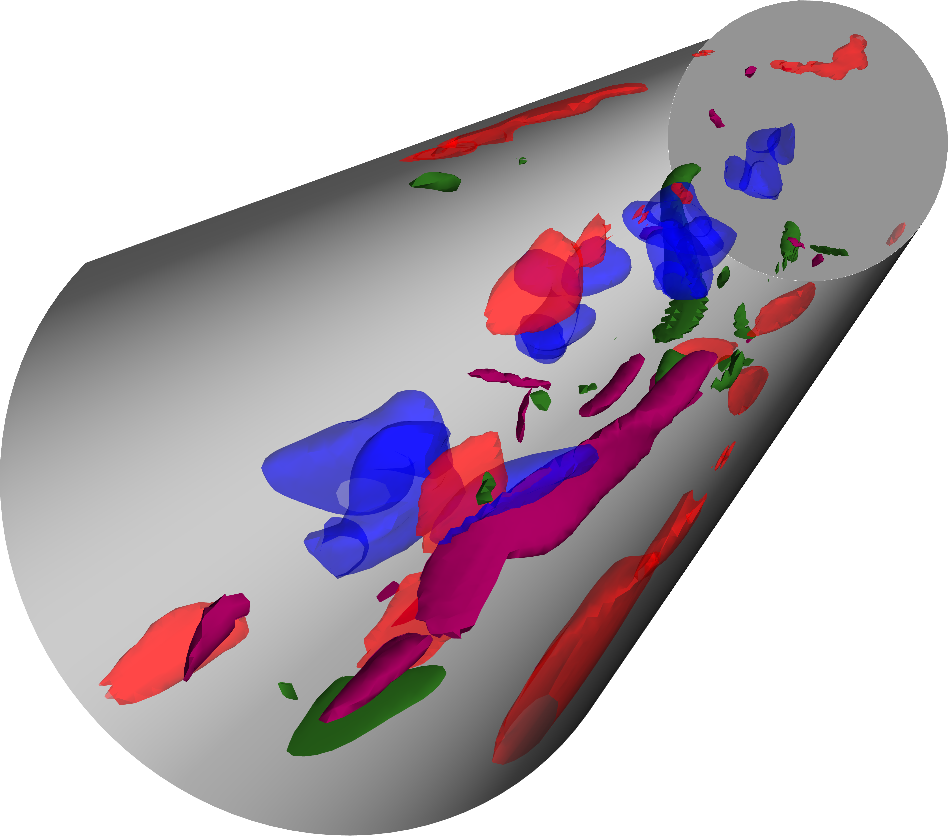} 
		\end{flushright}
	\end{minipage}
	\caption[capt]{
	(a) First approximation to the unstable manifold of \Sone , 
	as forward-integrated trajectories with initial conditions
	\refeq{e-S1unstManIC} visualized as projections onto the 
	local coordinate frame \refeq{e-projBases}. $L$ : Laminar 
	solution.
	Inset: Same $12$ trajectories on the unstable manifold for 
	time interval $\zeit \in [0, 2.5]$, illustrating the initial 
	almost uniform expansion.
	Isosurfaces of streamwise velocity and vorticity at 
	(b) $u = 0.36$ (red), $u = -0.50$ (blue),
	$\omega = \pm 0.99$ (green/purple), 
	(c) $u = 0.50$ (red), $u = -0.62$ (blue)
	$\omega = \pm 1.5$ (green/purple), 	
	(d) $u = 0.66$ (red), $u = -0.71$ (blue)		
	$\omega = \pm 9.0$ (green/purple).
	Flow direction into the page.  	
    \label{f-man1winst}
	}
\end{figure}

As a first approximation to \Sone's unstable manifold, we start 
trajectories with the initial conditions
\beq
    \sspRed_\phi (\zeit = 0) = \sspRed_{\Sone} + \epsilon 
                \left(\frac{\evecRed{\Sone}{1}}{\eval{\Sone}{1}} \cos \phi
                    + \frac{\evecRed{\Sone}{2}}{\eval{\Sone}{2}} \sin \phi \right) \, ,
\label{e-S1unstManIC}
\eeq
where $\sspRed_{\Sone}$ is the symmetry-reduced \statesp\ point 
corresponding to \Sone,  $\evecRed{\Sone}{1}$ and $\evecRed{\Sone}{2}$ 
are \Sone's
leading linear stability eigenvectors projected onto the slice as 
described in \refeq{e-Projz}, and $\epsilon = 10^{-4}$ is a small 
constant. \refeq{e-S1unstManIC} defines an ellipse on 
the $(\evecRed{\Sone}{1}, \evecRed{\Sone}{2})$-hyperplane, 
parametrized by $\phi$ and the
scaling of perturbations by corresponding eigenvalues $\eval{\Sone}{1,2}$
lets trajectories expand initially at similar rates 
\rfp{Krauskopf2005}. This is illustrated on the inset of 
\reffig{f-man1winst} (a) where 
$\sspRed_\phi (\zeit)$ for $\zeit \in [0, 2.5]$ is shown for $12$
equally spaced trajectories in $\phi \in [0, 2 \pi)$ as projections
onto the local bases formed by orthogonalizing 
$\evecRed{\Sone}{1}, \evecRed{\Sone}{2}$, \ie\ 
\beq
	e_1 = \inprod{\sspRed_\phi (\zeit) - \sspRed_{\Sone}}{
				  \evecRed{\Sone}{1, \perp}} \,, \quad
	e_2 = \inprod{\sspRed_\phi (\zeit) - \sspRed_{\Sone}}{
				  \evecRed{\Sone}{2, \perp}} \, .
	\label{e-projBases}
\eeq
In \refeq{e-projBases}, subscript $\perp$ indicates that 
$\evecRed{\Sone}{1}, \evecRed{\Sone}{2}$ are orthonormalized via 
the Gram-Schmidt 
procedure; 
\ie\ $\evecRed{\Sone}{2, \perp}$ is formed by subtracting 
$\evecRed{\Sone}{2}$'s projection onto the $\evecRed{\Sone}{1}$ 
direction and normalization. 
The inset of \reffig{f-man1winst} (a) illustrates that
these bases capture local dynamics very well. 
	In particular, 
	note that the trajectories starting from the initial conditions 
	\refeq{e-S1unstManIC}
	with $\phi = k \pi/2,\, k=0,1,2,3$ 
	are initially straight as they correspond to the perturbations 
	in the symmetry-reduced eigenvector directions; 
	whereas the rest appear bend since
	the two unstable directions expand at different rates. 
Outer panel of
\reffig{f-man1winst} shows these trajectories for longer times, 
until they either become turbulent or laminarize. The trajectories that
become turbulent or relaminarize are respectively colored pink and
gray. While the projections are locally reliable, they fail to 
fully capture the dynamics away from \Sone\ since the 
trajectories seem to
fold onto themselves. In order to illustrate the qualitative 
features of the transition, we marked $3$ instances on 
$\sspRed_\pi (\zeit)$ at $\zeit \in [40, 60, 90]$ and visualized
the corresponding three-dimensional flow structures on 
\reffig{f-man1winst} (b, c, d). Note that the structures on
\reffig{f-man1winst} (b) are quite similar to those of \Sone, while
the isosurface levels are set to higher values. This illustrates
the first part of the transition: As the trajectory moves from \Sone\
towards turbulence, streaks and vortices are amplified while their 
overall shape is more or less unchanged. As can be seen on 
\reffig{f-man1winst} (c), this initial amplification is followed by
the breakdown of streaks, and eventual turbulence (d) no longer 
exhibits structures that are coherent throughout the computational 
domain.
 
\begin{figure}[h]
	\centering
	(a) \includegraphics[height=0.43\textwidth]{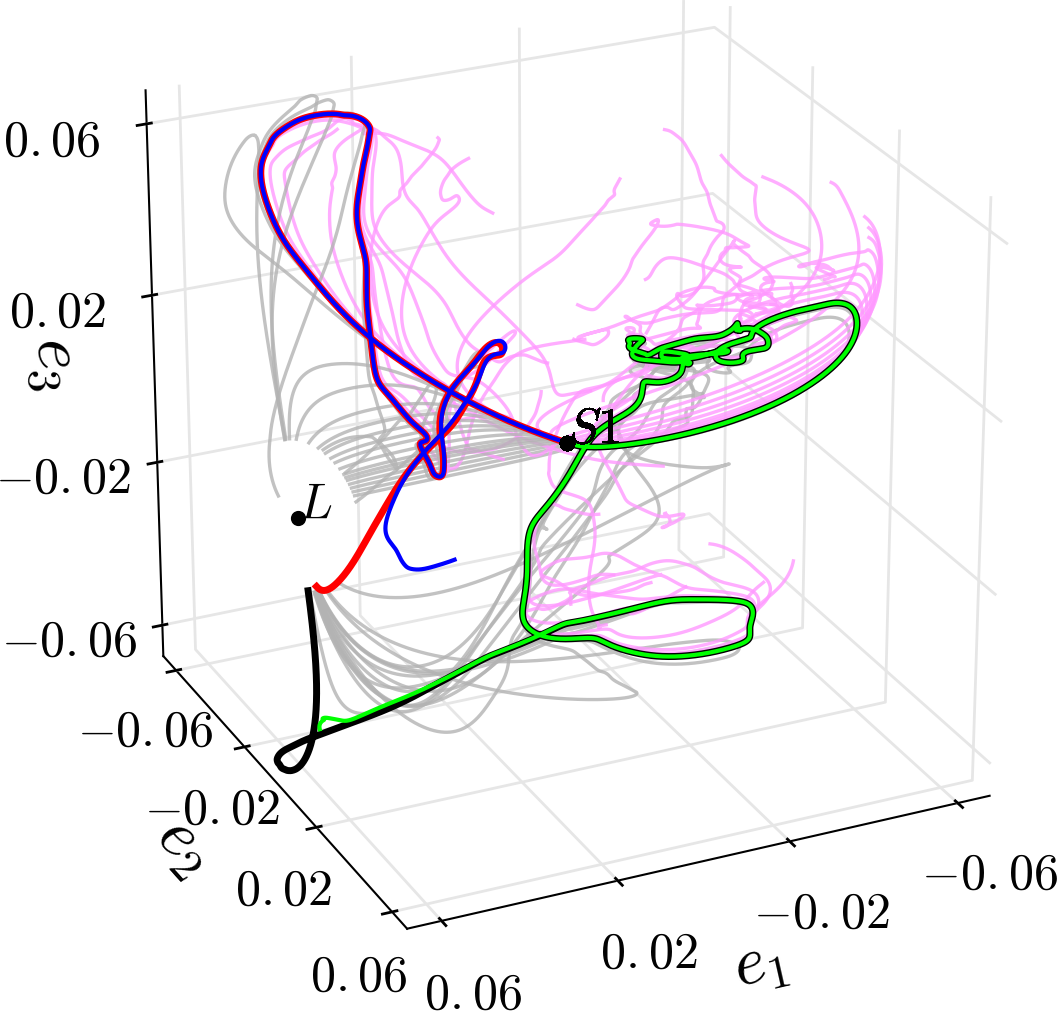} 
	(b) \includegraphics[height=0.43\textwidth]{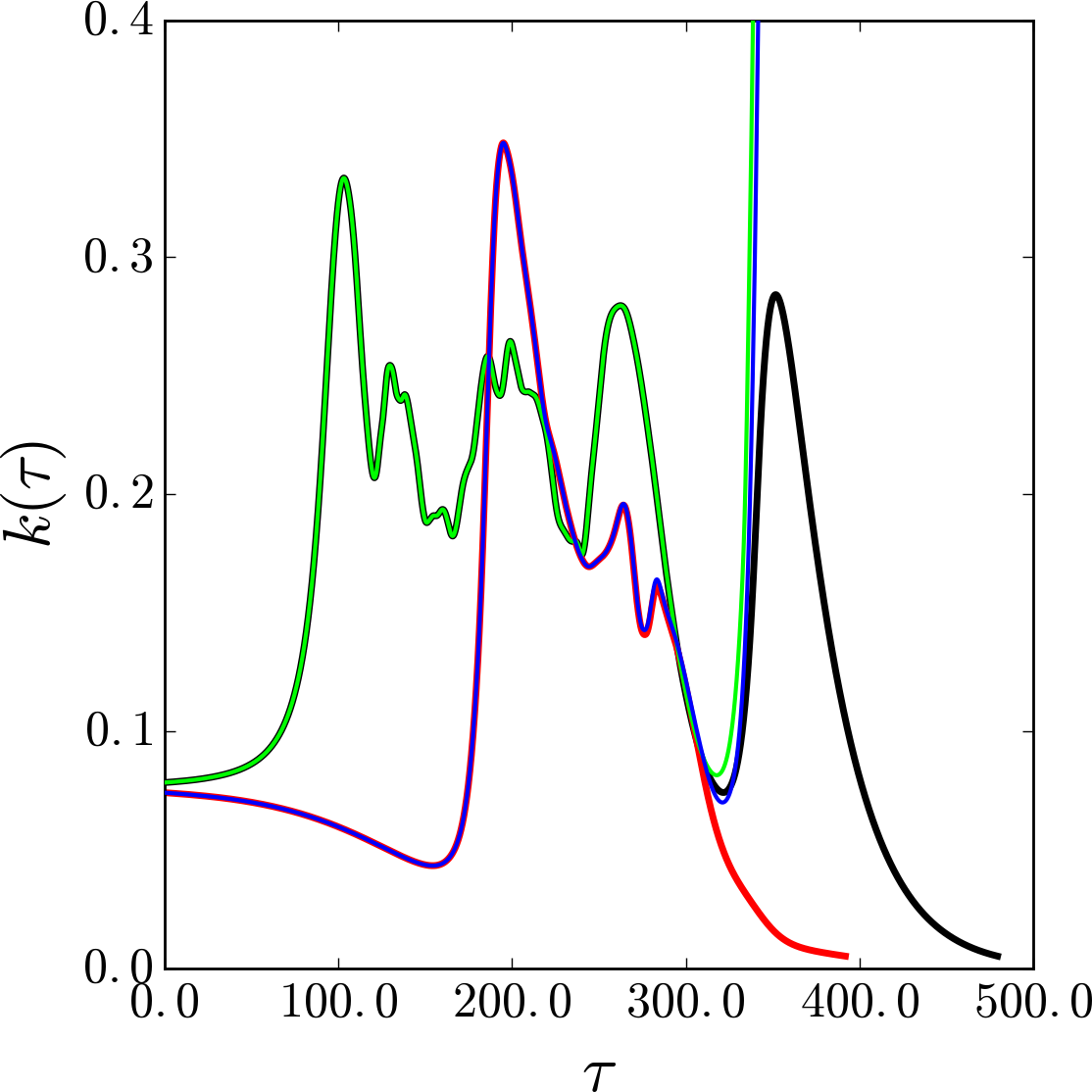} 
	\caption[]{
		(a) Unstable manifold of \Sone\ approximated by bisecting
		between trajectories that relaminarize and those that develop
		into turbulence. 
		(b) Time-series of kinetic energy for trajectories that stay
		on the edge for longest times.
		\label{f-man3DS1}
	}
\end{figure}

The $12$-trajectory approximation to the $2$-dimensional
unstable manifold is initially successful and illustrates the general
features of the unstable manifold, such as relaminarization and 
transition to turbulence. It is clear from \reffig{f-man1winst}(a) that 
this approximation very quickly fails to cover the extend of the 
manifold that stays in the laminar-turbulent boundary. In order to 
uncover details of this part, we bisected between trajectories that
transition to turbulence or relaminarize by changing $\phi$ in 
\refeq{e-S1unstManIC} until reaching the limit of numerical precision, 
as done in \rf{duguet07}. Differently, however,
we enforced shift-and-reflect invariance in the time-stepping, 
in order to avoid numerical errors that could take trajectories 
outside the invariant subspace. In order to improve visibility, we 
visualized the unstable manifold approximated this way 
in three dimensions, using the
least contracting stability eigenvector $\evecRed{\Sone}{4,\perp}$ 
(symmetry reduced and orthonormalized) in addition to the leading-two
\refeq{e-projBases} as the third basis. The unstable manifold 
visualized this way is shown in \reffig{f-man3DS1}(a), where we coloured 
trajectories that eventually laminarize gray, and those that 
transition to turbulence pink. Additionally, we plotted the last four 
trajectories in the bisection procedure thicker than the rest, 
using colors red, blue, green, and black. For comparison, time series
of the kinetic energy 
for these four trajectories is shown in 
\reffig{f-man3DS1}(b), where we see that the trajectories stay on the 
`edge' longer than $300$ time units. 

\begin{figure}[h]
	\begin{minipage}{0.43\textwidth}
		(a) \includegraphics[height=0.9\textwidth]{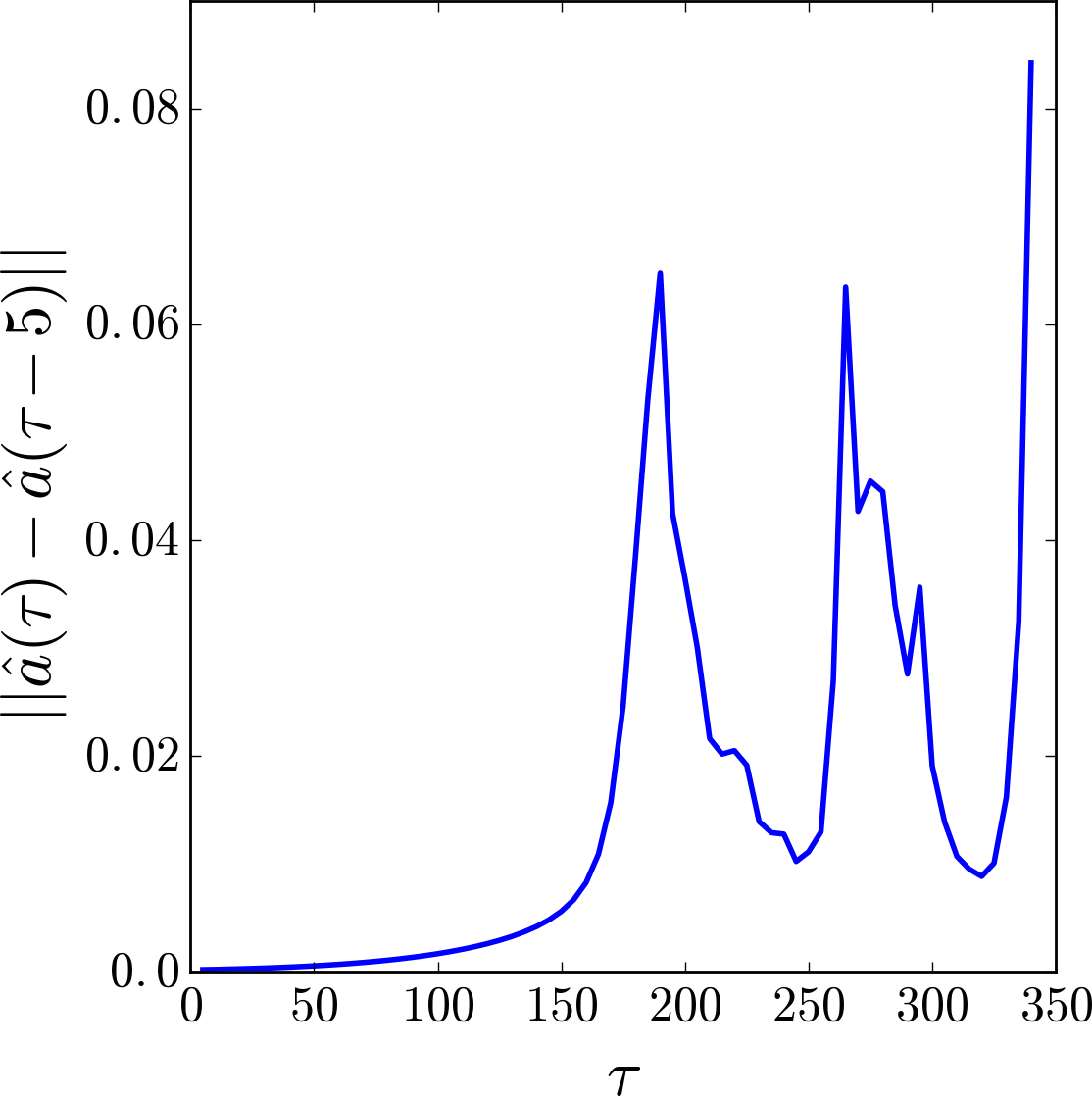}
	\end{minipage}
	\begin{minipage}{0.55\textwidth}
		\begin{flushright}
			(b) \includegraphics[height=0.35\textwidth]{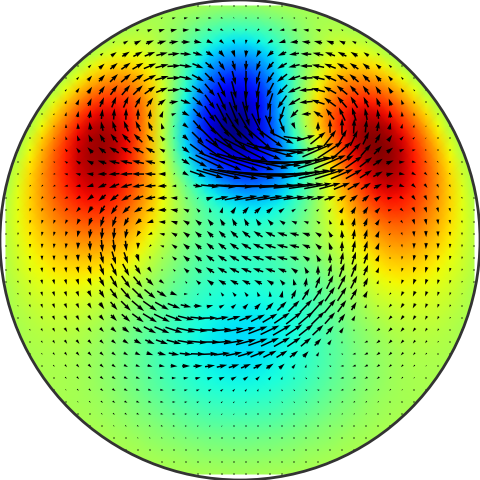} 
			(c) \includegraphics[height=0.35\textwidth]{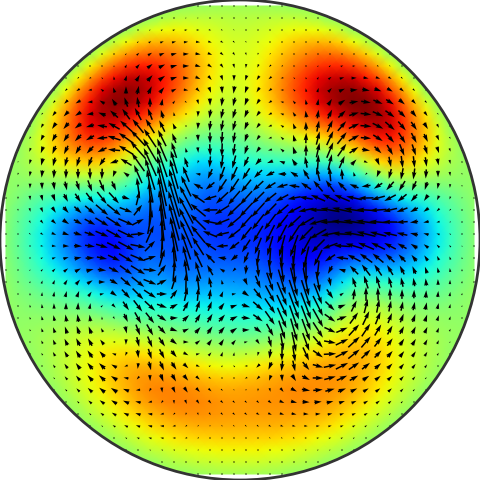} \\
			(d) \includegraphics[height=0.35\textwidth]{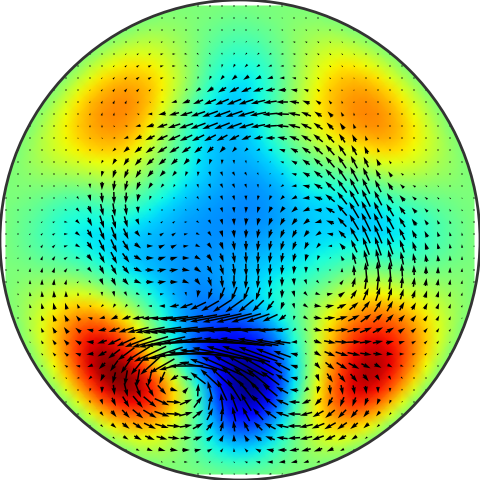} 
			(e) \includegraphics[height=0.35\textwidth]{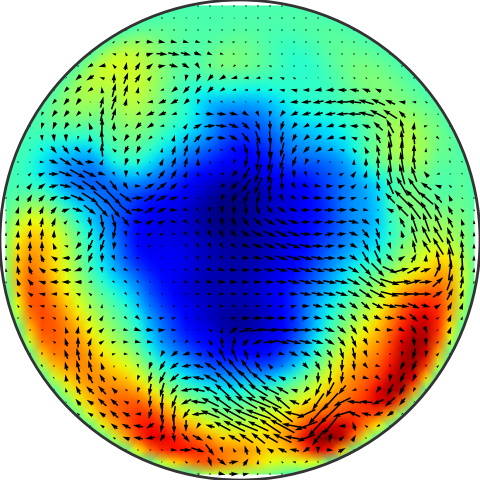} 
		\end{flushright}
	\end{minipage}
	\caption[capt]{
		\label{f-recurrence}
		(a) Self-recurrence measured over an edge trajectory 
		(drawn blue in \reffig{f-man3DS1}) on the unstable manifold
		of \Sone .
		Color-coded (fast/slow:red/blue) stream-wise velocity 
		and cross-sectional velocity (arrows) averaged over half
		pipe length ($z \in [0, L/2]$) at 
		(b) initial time $\zeit=0$, 
		(c) local minimum $\zeit=245$, 
		(d) local minimum $\zeit=320$,
		(e) final time $\zeit=345$.
		Flow direction into the page.
	} 
\end{figure}

Upon our investigation of the trajectories that stay on the edge for
the longest times, we observed episodes where \statesp\ trajectories slow 
down, which might indicate visits to the neighborhoods of other 
traveling waves. In order to quantify this, we measured the 
``self-recurrence'', which we defined as the norm of the difference 
between the symmetry-reduced states at time $\zeit$ and $\zeit - 5$ 
on the same trajectory. Note that in the symmetry-reduced \statesp , 
the traveling waves are equilibria thus no extra care for the 
translation symmetry is necessary.
\refFig{f-recurrence} shows the self-recurrence (a) over the 
trajectory which is shown blue on \reffig{f-man3DS1} and flow 
structures at different times on this trajectory (b-e).

As the trajectory evolves on the edge, it visits states with 
qualitatively different features: At the first minimum 
(\reffig{f-recurrence} (c)) of the 
self-recurrence function, new fast streaks begin to form on 
the down-side of the pipe cross-section and at the second minimum 
(\reffig{f-recurrence} (d)) initial fast streaks begin to disappear.
Eventual transition to turbulence (\reffig{f-man3DS1} (e)) has 
quantitative features similar to those illustrated in 
\reffig{f-man1winst}: streaks amplify, spread, and break into smaller
scale structures on the opposite side of the pipe. 
Newton-Krylov-hookstep searches starting from initial conditions 
visualized on \reffig{f-recurrence} (c) and (d) converged to 
traveling wave solutions: Starting from the latter, we find $\Sone$, 
rotated by $\pi$ about the pipe axis; whereas the Newton search 
starting from \reffig{f-recurrence} (c) lead us to a previously 
unknown traveling wave, which we study in the next section.

\subsection{Unstable manifold of \SoneN}
\label{s-SoneN}

\begin{figure}[h]
	\centering
	(a) \includegraphics[height=0.25\textwidth]{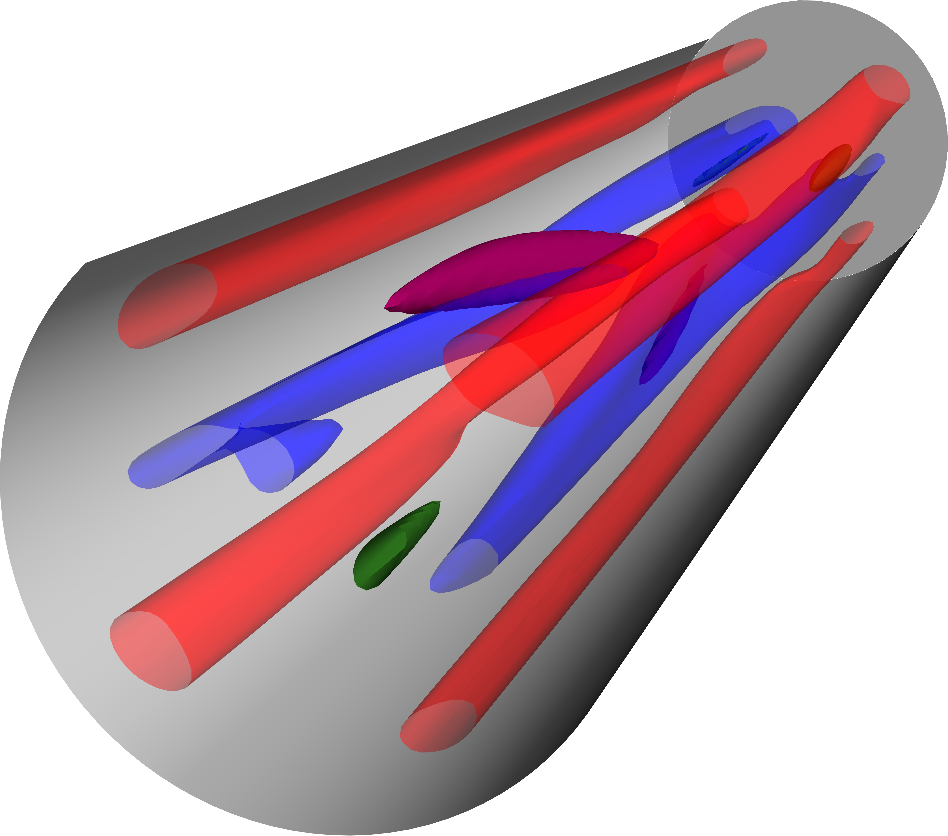} 
	(b) \includegraphics[height=0.25\textwidth]{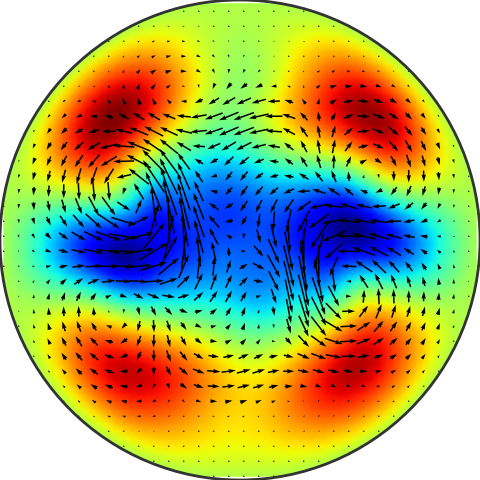} 
	(c) \includegraphics[height=0.25\textwidth]{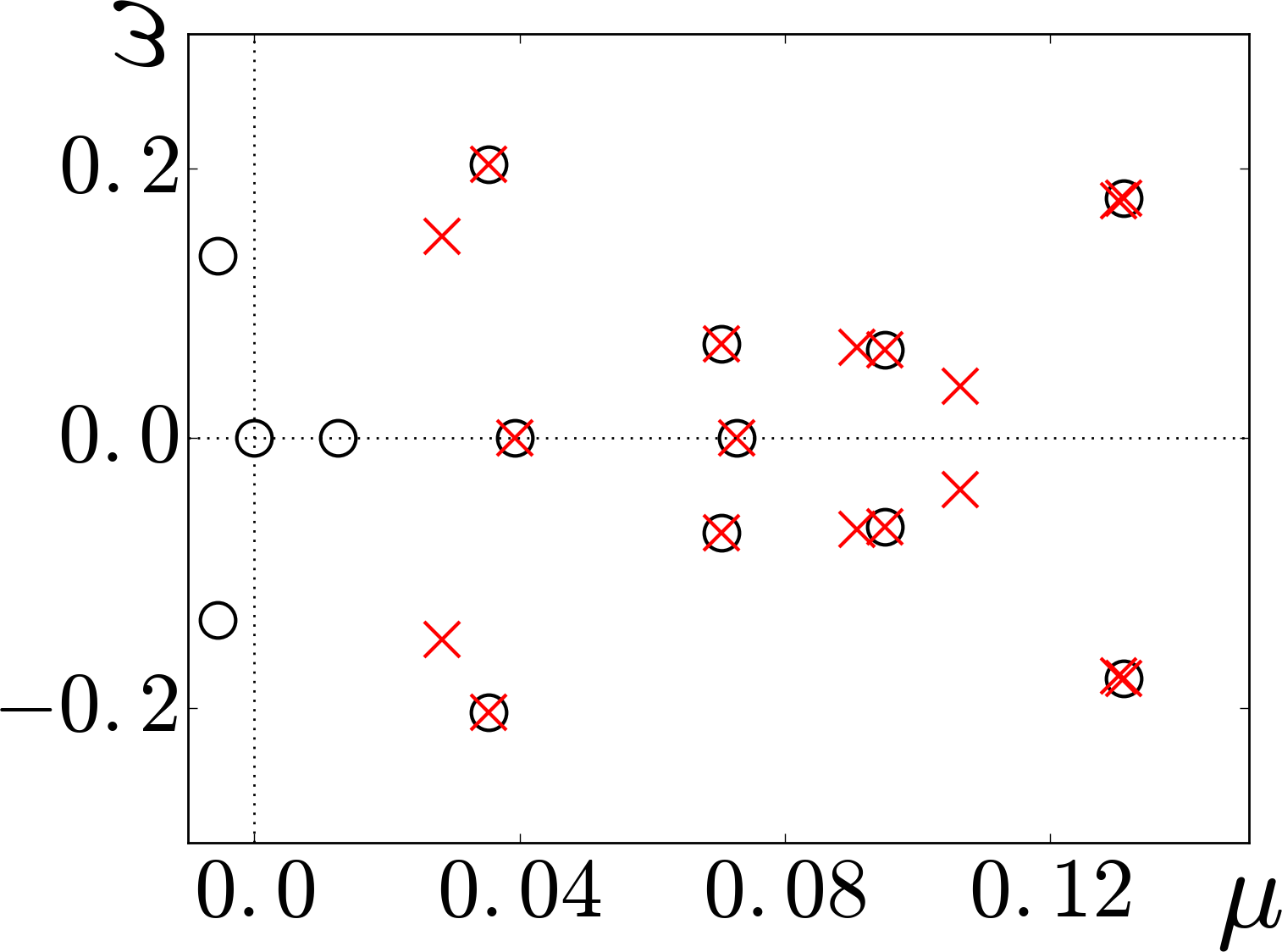} 
	\caption[]{
		(a) Isosurfaces of streamwise velocity at 
		u = 0.27 / -0.38 (red / blue) 
		and streamwise vorticity at 
		$\omega_z = \pm 0.96$ (green / purple) of \SoneN.
		(b) 
		Color-coded (fast/slow:red/blue) streamwise velocity 
		and cross-sectional velocity (arrows) averaged over half
		pipe-length ($z \in [0, L/2]$) visualization of \SoneN .
		(c) Leading linear stability eigenvalues of \SoneN\ 
		on complex plane computed in shift-and-reflect invariant 
		subspace (black circles) and full \statesp\ (red crosses).
		\label{f-s1n}
	}
\end{figure}

\refFig{f-s1n}(a,b) shows three- and two-dimensional 
visualizations of the new solution, which we hereafter refer to as 
$\SoneN$.
While being on the laminar-turbulent boundary, \SoneN\ has a much 
more complicated structure compared to \Sone ; its kinetic energy is
roughly twice of it; and it has many more unstable directions. 
Solutions with similar properties were known 
\rfp{FE03,WK04,duguet07,Pringle09}, 
mostly with higher azimuthal symmetries. However, none of the 
previously known solutions' relevance to the full problem 
(without imposed symmetries) were established.
Similarities between the initial condition's 
\reffig{f-recurrence} (c) and the converged state's flow structures 
\reffig{f-s1n} (b) are apparent. Note particularly the 
locations of the vortical structures, which align very well. 
Since these structures do not extend along 
the pipe axis like streaks do, their one-to-one comparison is made
possible by reduction to the translation-reduced \statesp\ 
\refeq{e-sspred}. 

\begin{figure}[h]
	\centering
	(a) \includegraphics[height=0.27\textwidth]{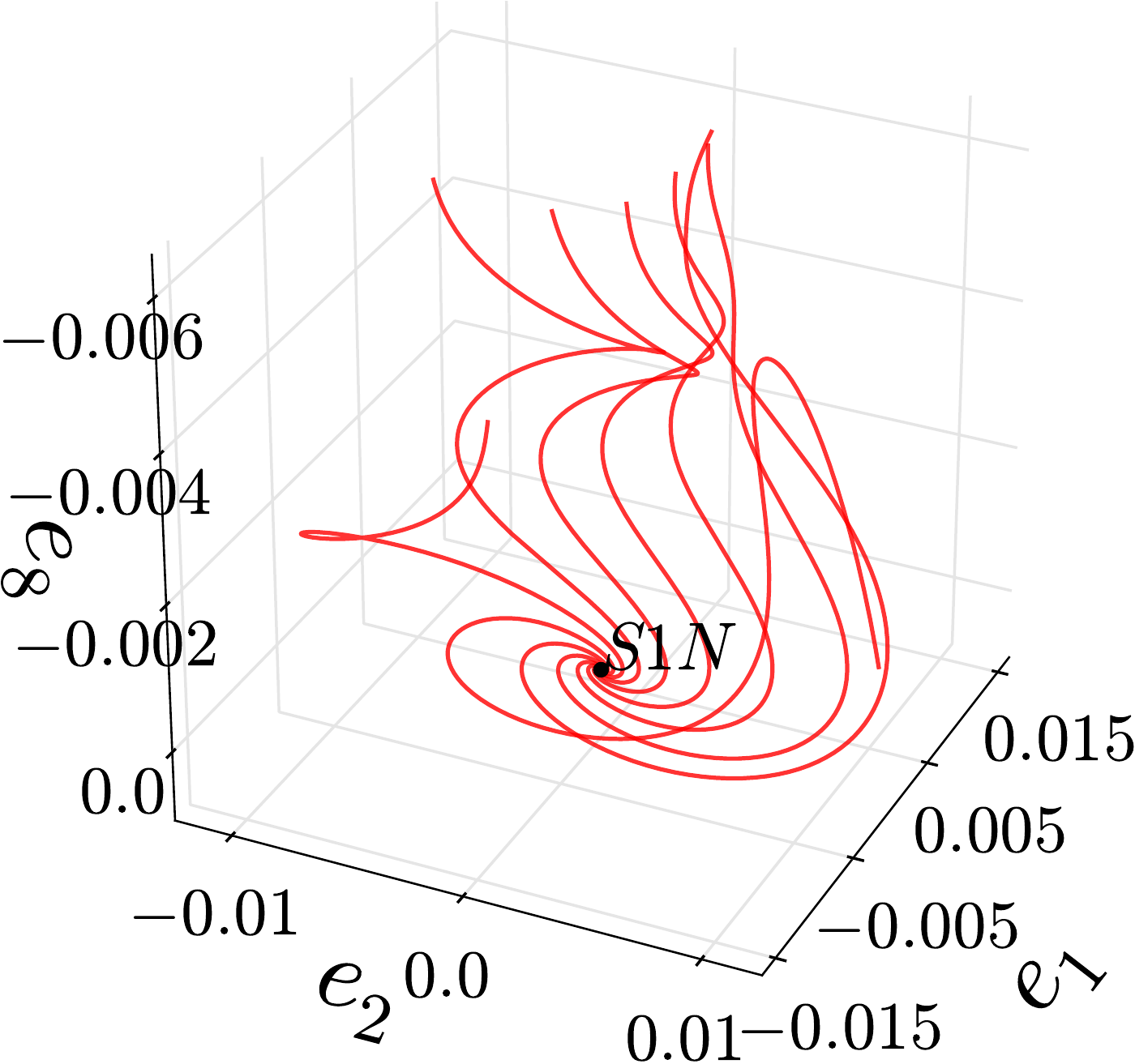} 
	(b) \includegraphics[height=0.27\textwidth]{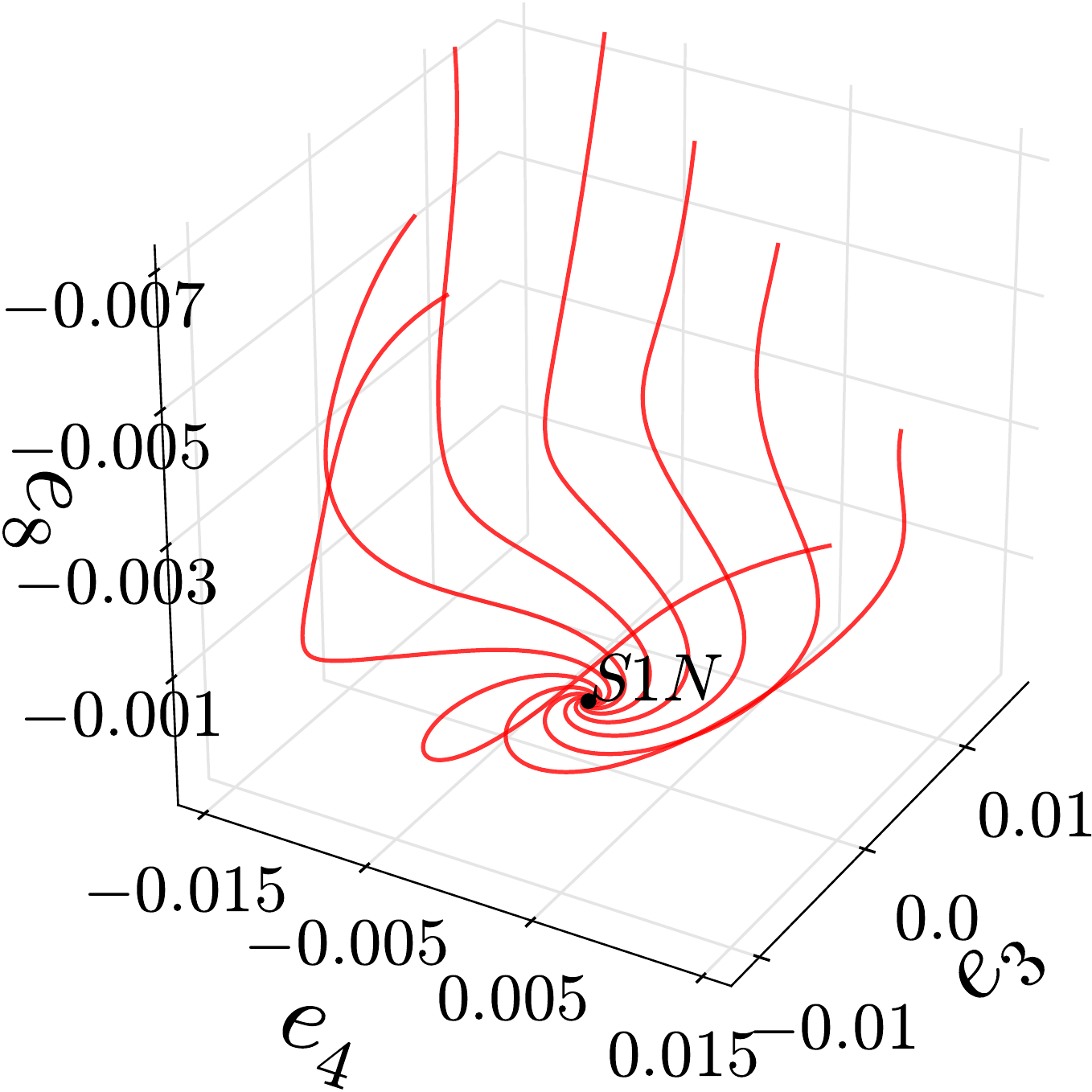} 
	(c) \includegraphics[height=0.27\textwidth]{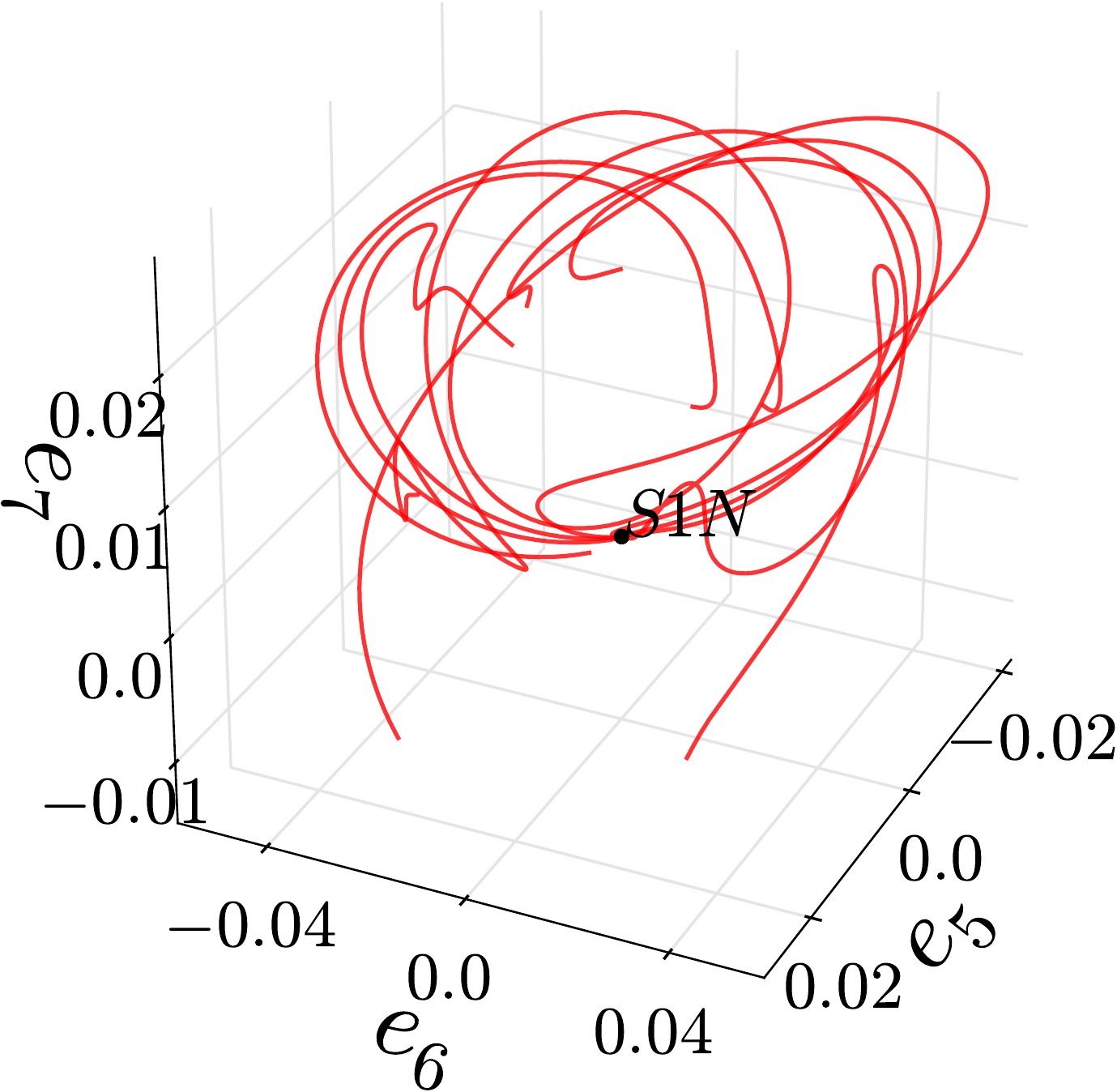} 
	\caption[]{
		Two-dimensional surfaces associated with leading 
		three unstable eigenvector of \SoneN\ approximated 
		as set of trajectories with initial conditions given
		by \refeq{e-SpiralOut} for $k = 1,3,5$ (a, b, c).
		\label{f-s1nManifolds}
	}
\end{figure}

As shown in \reffig{f-s1n} (c), \SoneN's unstable manifold is 
11-dimensional in the shift-and-reflect subspace and more unstable
eigenvalues were found when the symmetry restriction was lifted.
This renders a complete computational study of its unstable manifold
impractical. Therefore on \reffig{f-s1nManifolds}, 
we visualize two-dimensional 
surfaces associated with each of 
the leading three complex unstable eigenvalues
in the full \statesp\ as the first application of symmetry 
reduction in both $z$ and $\theta$.

Let $\eval{\SoneN}{k} = \mu^{\SoneN}_k + i \omega^{\SoneN}_k$ 
and $\evec{\SoneN}{k} = U^{\SoneN}_k + i W^{\SoneN}_k$ 
respectively be complex eigenvalues and eigenvectors of \SoneN. 
Set of trajectories that approximately covers the linearized dynamics
in the local two-dimensional plane spanned by 
$(U^{\SoneN}_k, W^{\SoneN}_k)$ are 
given by
\beq
\sspRRed_\delta (\zeit = 0) = \sspRRed_{\SoneN} 
+ \epsilon \exp (2 \pi \mu^{\SoneN}_k \delta / \omega^{\SoneN}_k)
\tilde{U}^{\SoneN}_k\,,
\mbox{where }
\delta \in [0, 1) \, .
\label{e-SpiralOut}
\eeq
In \refeq{e-SpiralOut}, $\epsilon$ is a small number and  
$\tilde{} $ implies that \statesp\ points and 
eigenvectors are transformed into the \fFslice\ via \refeq{e-ssprred}
and \refeq{e-Projtheta} respectively. Under the linearized dynamics, 
a small perturbation to $\sspRRed_{\SoneN}$ in $\tilde{U}_k$ 
direction expands by 
$\exp(2 \pi \mu^{\SoneN}_k / \omega^{\SoneN}_k)$ at one return of 
the spiral. Thus, the trajectories starting from the initial 
conditions  \refeq{e-SpiralOut} approximately satisfy 
$\sspRRed_{\delta = 0} (2 \pi / \omega^{\SoneN}_k) 
= \sspRRed_{\delta = 1} (0)$, covering the associated 
two-dimensional surface. Using this approximation, we visualized
time-forward dynamics of $2D$-surfaces associated with 
$\evec{\SoneN}{1,2}$, 
$\evec{\SoneN}{3,4}$, and 
$\evec{\SoneN}{5,6}$ on \reffig{f-s1nManifolds}
(a, b, c) respectively. We assume that the eigenvalues are ordered
from most unstable to the least \ie\  
$\Real \eval{\SoneN}{1} \geq 
 \Real \eval{\SoneN}{2} \geq 
 \Real \eval{\SoneN}{3} \geq \ldots$ 
Note that complex eigenvalues and eigenvectors satisfy
$\eval{\SoneN}{1} = \eval{\SoneN*}{2}$, 
$\eval{\SoneN}{3} = \eval{\SoneN*}{4}$, 
$\eval{\SoneN}{5} = \eval{\SoneN*}{6}$, 
$\evec{\SoneN}{1} = \evec{\SoneN*}{2}$, 
$\evec{\SoneN}{3} = \evec{\SoneN*}{4}$, 
$\evec{\SoneN}{5} = \evec{\SoneN*}{6}$, 
where $*$ stands for complex conjugation. Projection bases 
${\bf e}_{1,2}$, ${\bf e}_{3,4}$, ${\bf e}_{5,6}$ are
generated from 
$\evecRRed{\SoneN}{1}$, 
$\evecRRed{\SoneN}{3}$, and 
$\evecRRed{\SoneN}{5}$ 
as follows:
By definition, 
$\evecRRed{'\SoneN}{k} = e^{i \phi} \evecRRed{\SoneN}{k}$ 
is a stability
eigenvector with eigenvalue $\eval{\SoneN}{k}$. 
If we choose 
\beq
	\phi = \frac{1}{2} \arctan{\frac{2 
			\inprod{\tilde{U}^{\SoneN}_k}{\tilde{W}^{\SoneN}_k}}{
			||\tilde{W}^{\SoneN}_k||^2 - ||\tilde{U}^{\SoneN}_k||^2}} \, , 
\eeq
$\tilde{U}^{'\SoneN}_k = \Real \evecRRed{'\SoneN}{k}$ and
$\tilde{W}^{'\SoneN}_k = \Imag \evecRRed{'\SoneN}{k}$ become orthogonal \ie\ 
$\inprod{\tilde{U}^{'\SoneN}_k}{\tilde{W}^{'\SoneN}_k} = 0$. 
Projection bases are formed from these vectors as 
$({\bf e}_{k}, {\bf e}_{k + 1}) = 
 (\tilde{U}^{'\SoneN}_k  / 
 ||\tilde{U}^{'\SoneN}_k ||, 
 \tilde{W}^{'\SoneN}_k / ||\tilde{W}^{'\SoneN}_k||)$
 for $k = 1, 3, 5$.
These bases fully capture local two-dimensional dynamics. We chose
the third projection direction by trial and error to capture as much 
as possible as the trajectories develop into turbulence: 
${\bf e}_7 = \evecRRed{\SoneN}{11} / ||\evecRRed{\SoneN}{11}||$ 
and 
${\bf e}_8 = \evecRRed{\Sone}{1} / ||\evecRRed{\Sone}{1}||$.
$\evecRRed{\SoneN}{11}$ is the real unstable eigenvector of 
$\SoneN$ with largest 
real eigenvalue $\eval{\SoneN}{11}$ and $\evecRRed{\Sone}{1}$ 
is the leading 
eigenvector of $\Sone$. We have already illustrated in 
\refsect{s-SoneN} that the $\evec{\Sone}{1}$ direction corresponds to the
trajectories in the vicinity of $\Sone$, which either laminarize 
or become turbulent and we are going to demonstrate that 
$\evec{\SoneN}{11}$ 
takes the same role for \SoneN .

\texttt{Openpipeflow} normalizes stability eigenvectors such that
their norms are equal to that of the associated solution, \ie\ 
$||\evec{TW}{k}|| = || \ssp_{TW} ||$. For computations of 
\reffig{f-s1nManifolds}, we set 
$\epsilon = 10^{-4}$ (a, b)
and $\epsilon = 10^{-8}$ (c); and picked $8$ equidistant point in $[0, 1)$
for $\delta$. While the spiral-out dynamics is 
clearly visible on \reffig{f-s1nManifolds} (a) and (b), trajectories
look less organized on panel (c). This is because the imaginary part 
of the fifth eigenvalue $\eval{\SoneN}{5} = 0.106 + i 0.0383$ is rather 
small, rendering the time for the trajectories to complete a full
spiral long. This is not the case for 
$\eval{\SoneN}{1} = 0.131 + i 0.178$ and
$\eval{\SoneN}{3} = 0.130 + i 0.176$. While we might have obtained a 
better representation of the $2D$ manifold in \reffig{f-s1nManifolds}(c)
if we had used more trajectories to represent it, we chose to leave it
as it is in order to illustrate the possible shortcomings of the
method. All trajectories in \reffig{f-s1nManifolds} eventually become 
turbulent, illustrating the rich dynamics that the laminar-turbulent 
boundary can exhibit.

\begin{figure}[h]
	\begin{minipage}{0.41\textwidth}
		    (a) \includegraphics[width=0.91\textwidth]{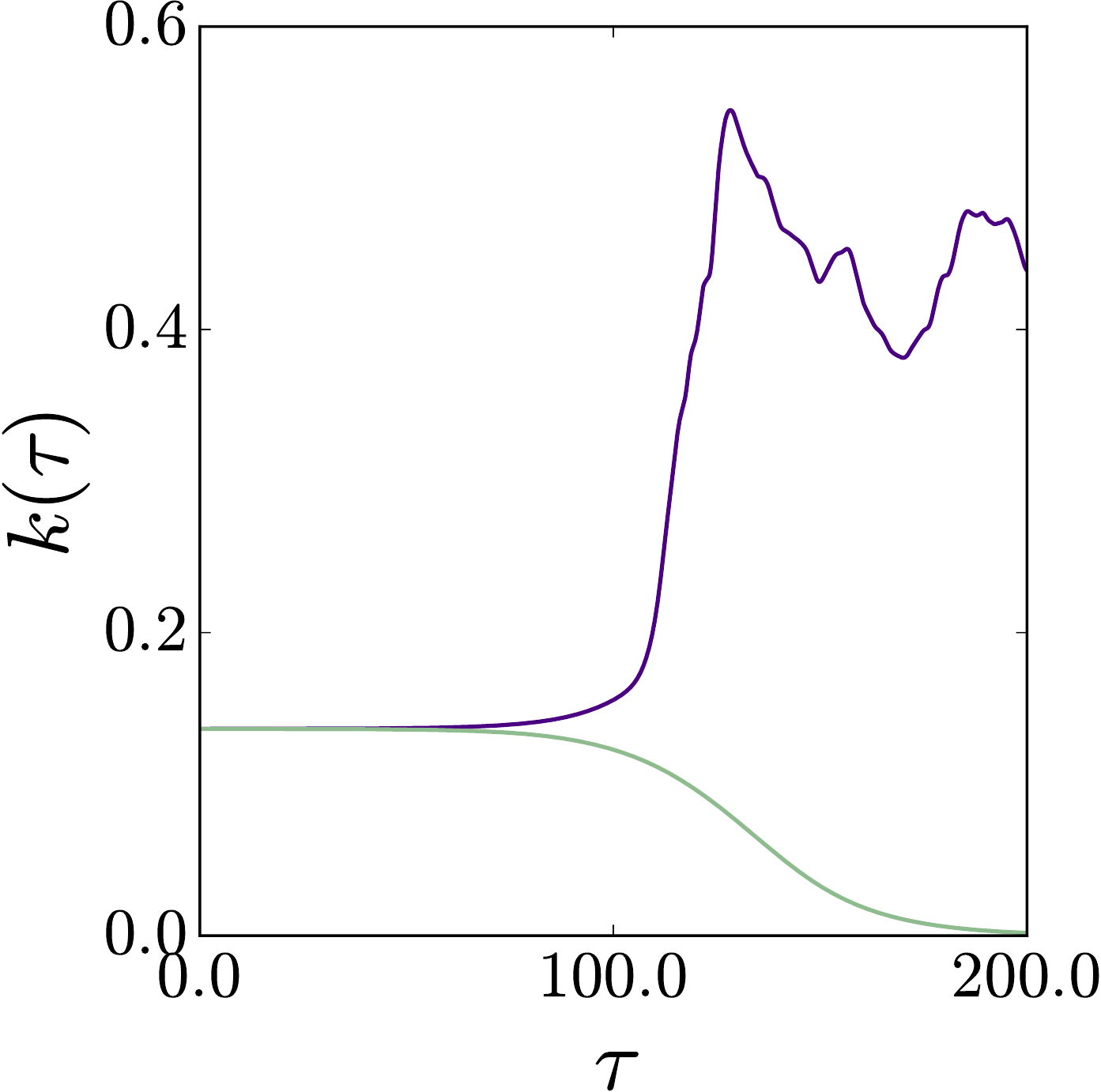} 
	\end{minipage}
	\begin{minipage}{0.58\textwidth}
		\begin{flushright}
			(b) \includegraphics[height=0.32\textwidth]{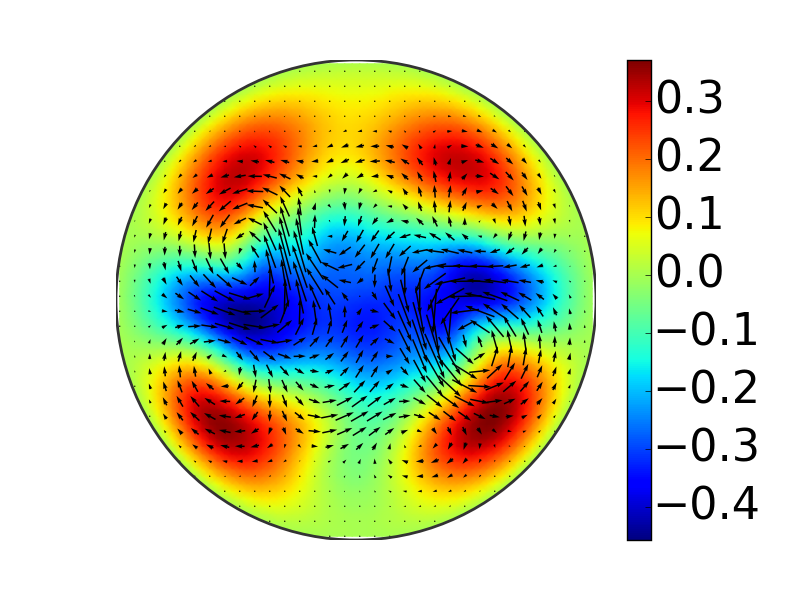} 
			(c) \includegraphics[height=0.32\textwidth]{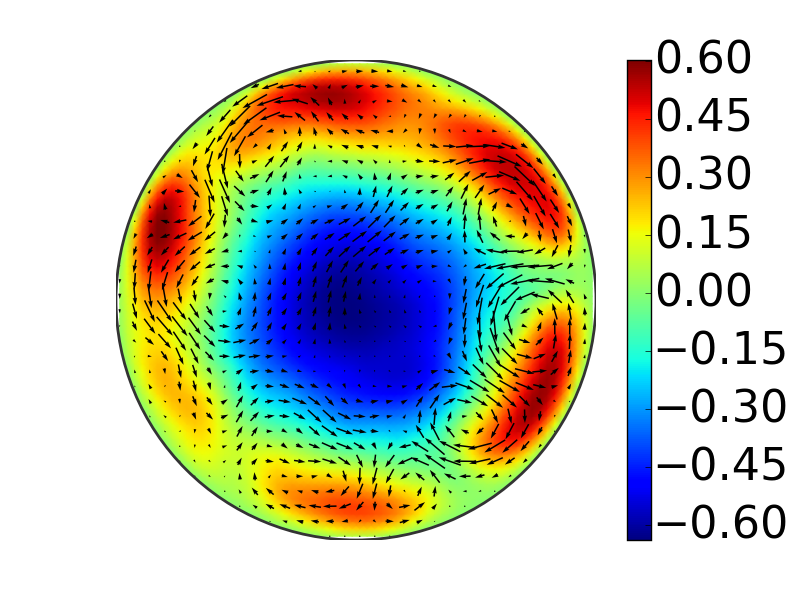} \\
			(d) \includegraphics[height=0.32\textwidth]{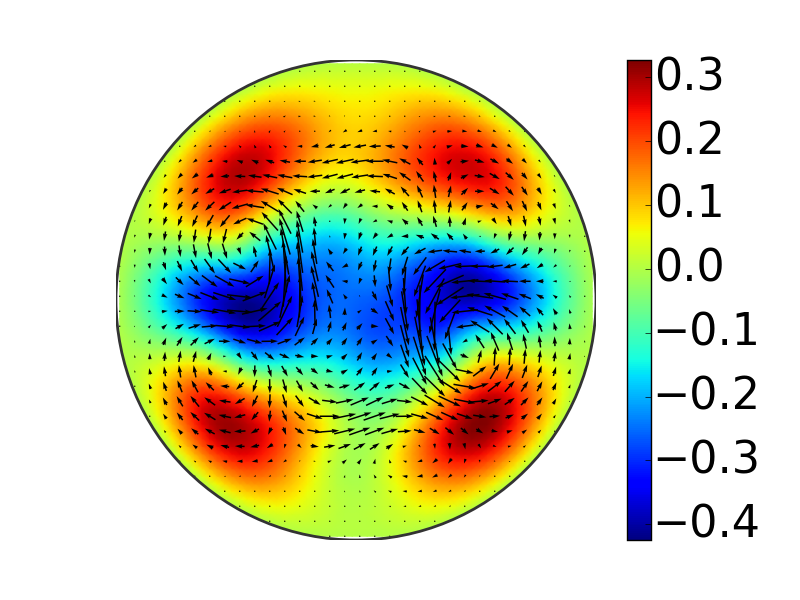} 
			(e) \includegraphics[height=0.32\textwidth]{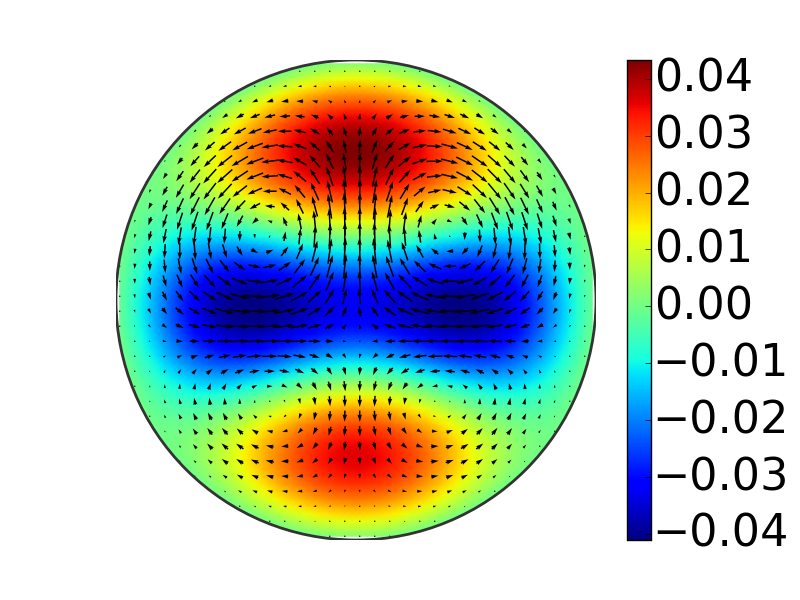} 
		\end{flushright}
	\end{minipage}
	\caption[capt]{
		\label{f-ekinperturbpm}
		(a) Time evolution of kinetic energy for trajectories starting
		from $\ssp_{\SoneN} \pm 10^{-4} \evec{\SoneN}{11}$ (green/purple).
		(b, c, d, e)
		Color-coded stream-wise velocity and cross-sectional velocity 
		(arrows) averaged over half pipe-length 
		($z \in [0, L/2]$) at times $\zeit = 100, 200$ respectively 
		on left and right.
		(b, c) Transition to turbulence (purple on panel a).
		(d, e) Laminarization (green on panel a).
	} 
\end{figure}

In the full \statesp , the leading $10$ stability eigenvalues of 
\SoneN\ are complex conjugate (\reffig{f-s1n} (c)) and 
$\eval{\SoneN}{11} = 0.0727$ 
is purely real. In \refFig{f-ekinperturbpm} (a), we show the 
time-evolution of the kinetic energy when \SoneN\ is perturbed in the
$\pm \evec{\SoneN}{11}$ direction. As one would expect from a solution on the 
laminar-turbulent boundary, the flow relaminarizes in one direction while 
becoming turbulent in the opposite. 
While this observation clearly shows that 
\SoneN\ also belongs to the laminar-turbulent boundary, whether or not 
if its presence influences generic edge trajectories is not known. In the
next section, we investigate this through a numerical experiment.

\subsection{Approaches to the traveling waves}
\label{s-numexp}

\begin{figure}[h]
	\setlength{\unitlength}{0.45\textwidth}
	(a) \includegraphics[width=\unitlength]{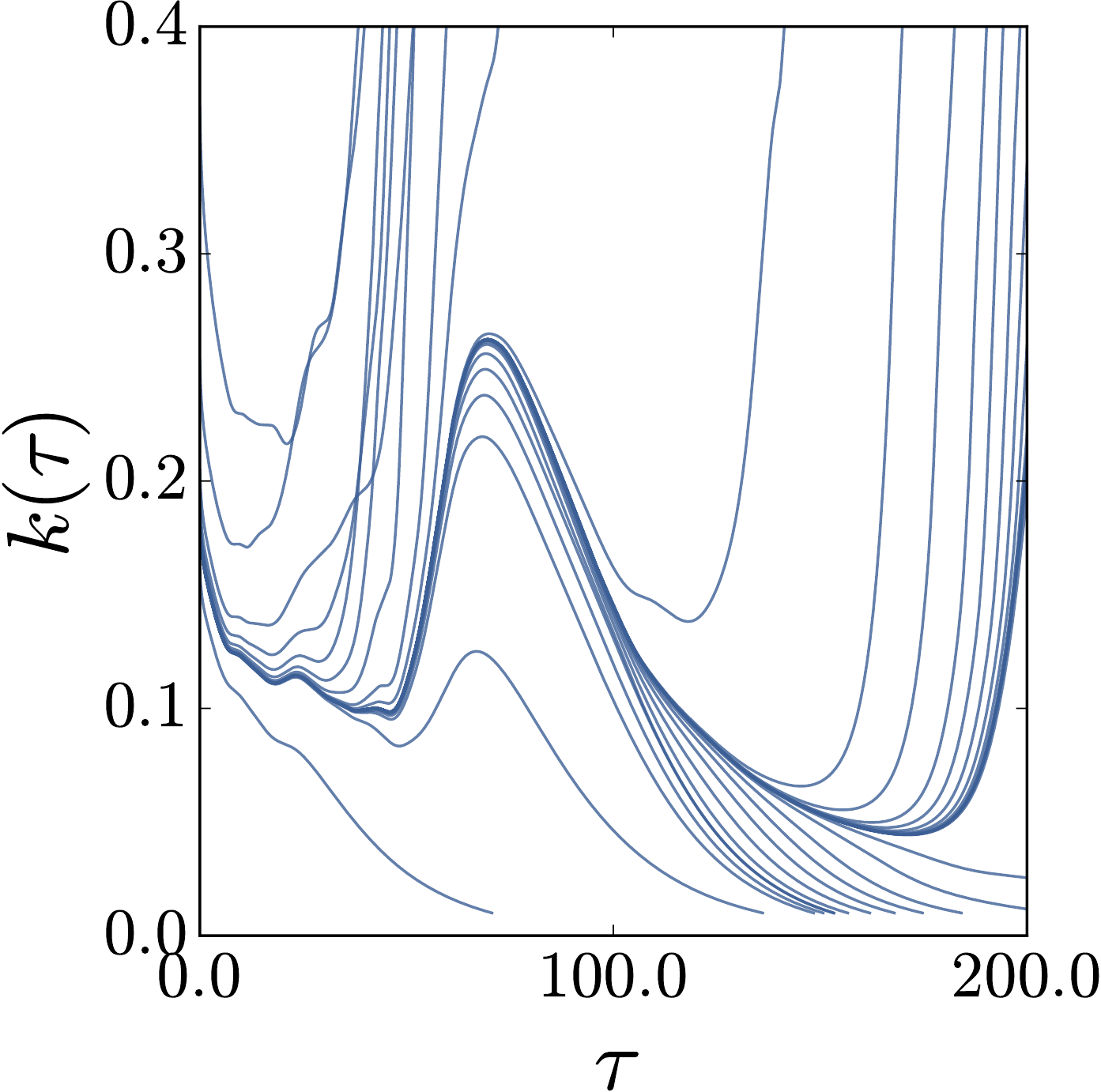}
	(b) \includegraphics[width=\unitlength]{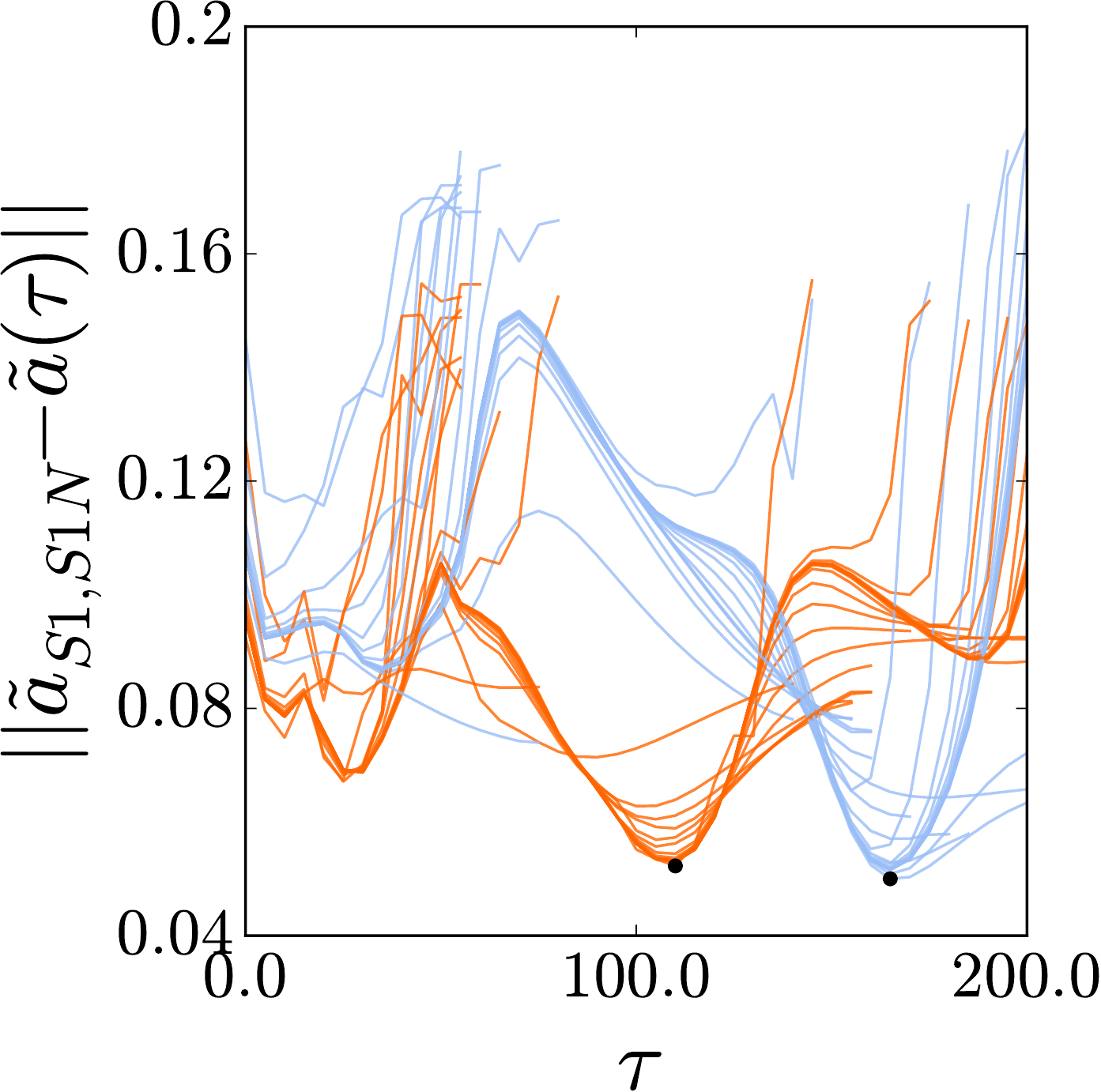}
	\caption{
		(a) Time evolution of turbulent kinetic energy for edge tracking 
		trajectories obtained through bisection. 
		(b) Distance of bisection trajectories from \Sone\ (blue) and \SoneN\ 
		(orange). Minima of distances at $\zeit = 110$ 
		($\min ||\sspRRed_{\SoneN} - \sspRRed (\zeit)|| $) and 
		at 
		$\zeit = 165$ ($\min ||\sspRRed_{\Sone} - \sspRRed (\zeit)|| $)
		are marked black.
		\label{f-edgeLower}}
\end{figure}

In order to illustrate how a generic trajectory on the 
laminar-turbulent is influenced by the traveling wave solutions 
present in the edge state, we carried out an
edge tracking\rf{IT01,SYE05} experiment where we bisected between
initial conditions that become turbulent and those that laminarize.
For this purpose, we randomly took a typical turbulent state at 
$\Reynolds = 10000$, scaled this field down by constants 
$c=0.5$ and $c=0.75$ and used these states as initial
conditions at $\Reynolds = 3000$. After observing that the former 
uneventfully proceeds towards the laminar solution while the latter 
triggers turbulence, 
we began generating new initial conditions by bisecting in $c$ until 
we reached the limit of our numerical precision, such 
that the resulting trajectories stay in the laminar-turbulent 
boundary for longer and longer times. Kinetic energy time-series of
these trajectories are shown in \reffig{f-edgeLower} (a). Such 
initial conditions are expected to approach to the invariant edge 
state\rf{SchEckYor07} irrespective of how the very first state is 
generated. Therefore our choice of initial state from 
$\Reynolds = 10000$ is arbitrary and many other initial states would
approach to the edge state following the same algorithm. 

\begin{figure}[h]
	\setlength{\unitlength}{0.2\textwidth}
	(a) \includegraphics[width=\unitlength]{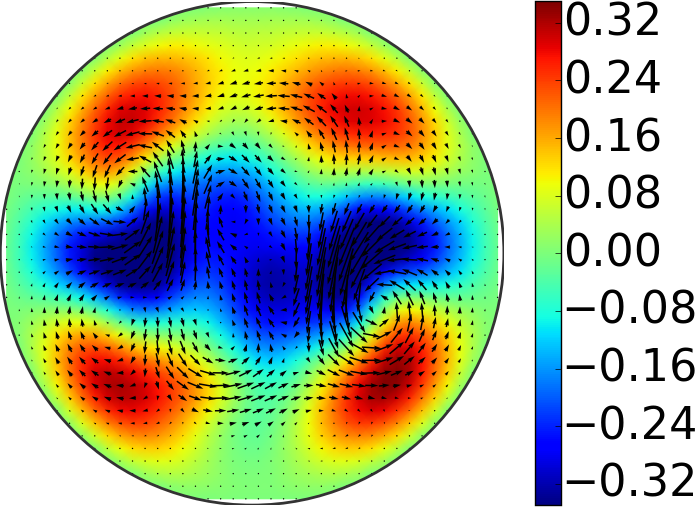}\,
	(b) \includegraphics[width=\unitlength]{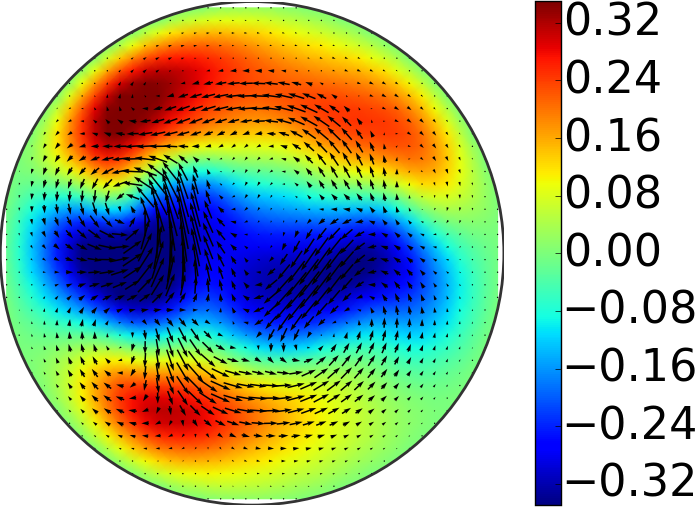}\,
	(c) \includegraphics[width=\unitlength]{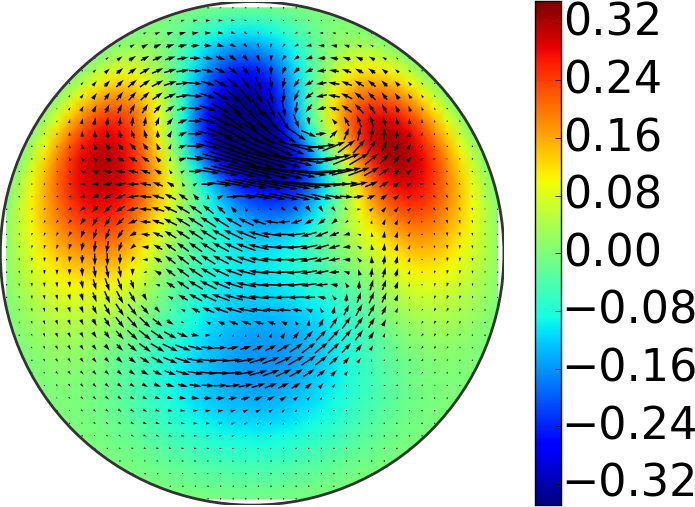}\,
	(d) \includegraphics[width=\unitlength]{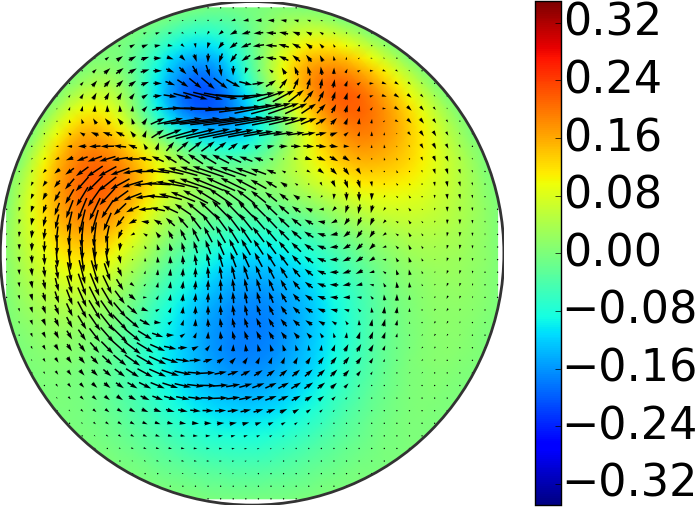}
	\caption{Color-coded streamwise velocity and cross-sectional 
		velocity (arrows) averaged over half-pipe length 
		($z \in [0, L/2]$) of
		(a) $\SoneN$, 
		(b) closest approach of the trajectory on the edge to $\SoneN$ at 
		$\zeit = 110$, 
		(c) $\Sone$, 
		(d) closest approach of the trajectory on the edge to $\Sone$ at 
		$\zeit = 165$. Each figure uses a fixed color-scale 
		$(\min, \max) = (-0.35, 0.35)$.
		\label{f-minDistLower}}
\end{figure}

For every trajectory we generated through edge-tracking we the sampled 
flow states at intervals of $\Delta \zeit = 5$ and brought these 
states to the \fFslice\ through \refeq{e-ssprred}. We then measured
distances of these states from \Sone\ and \SoneN\ within the \fFslice. 
$L_2$-distances of edge tracking trajectories from \Sone\ and \SoneN\ 
are shown blue and orange respectively in \reffig{f-edgeLower}(b), where
each set of curves have a clear minimum marked with a black dot. We 
visualized these closest approaches next to the traveling waves they 
approach to in \reffig{f-minDistLower}. Similarities between the flow
structures of the traveling waves and the edge trajectories near them 
are clear, although the correspondence is not one-to-one. 

Notice that without the symmetry-reduction, such an analysis would 
have required an optimization over axial translations and azimuthal 
rotations for each step in order to minimize the distance between two
states. Symmetry-reduction eliminates this step, decreasing the 
computational cost of the analysis tremendously.

\section{Conclusion and outlook}
\label{s:discus}
In this paper, we introduced a representation of the pipe flow, where
the continuous symmetries in axial and azimuthal directions are 
simultaneously reduced. This technical step was a straightforward 
extension of the \fFslice\ implementation of \refref{BudHof17}.
Nevertheless, it was not implemented before and it successfully 
closes the continuous symmetry reduction problem for pipe flow. 
Adapting this method to other canonical shear flows is 
straightforward. For instance, in a channel geometry, the role of 
axial and azimuthal coordinates are taken over by streamwise and 
spanwise coordinates; and one should only make a choice for the 
wall-normal dependence of the template functions. Since this choice 
is somewhat arbitrary, a convenient option could be
the first Chebyshev polynomial that is often used for numerical 
discretization in this direction. 

For the application, we decided to revisit the 
laminar-turbulent boundary in a short ($L\approx5$) periodic 
computational cell of the pipe flow. Our first calculation that 
approximated the unstable manifold of \Sone\ was very similar to that 
of Duguet \etal \rf{duguet07}, who conjectured that the unstable 
manifold of \Sone\ reached the neighborhoods of its own azimuthally 
rotated copies when it is followed along the laminar-turbulent 
boundary. Differently from \refref{duguet07}, we restricted our 
computation of the unstable manifold into the shift-and-reflect 
invariant subspace, to which the unstable manifold belongs. In 
addition, we visualized the unstable manifold as low-dimensional
projections from the \fFslice\ \refeq{e-sspred}. We found numerical
evidence that the portion of this unstable manifold that is confined
in the laminar-turbulent boundary visits the neighborhood of a 
new traveling wave, which we named \SoneN. Linear stability 
spectrum \reffig{f-s1n}(c) showed that \SoneN\ has a very high 
dimensional unstable manifold, albeit being on the laminar-turbulent 
boundary. By visualizing two-dimensional surfaces associated 
with leading three complex conjugate unstable eigenvectors on 
\reffig{f-s1n}, we illustrated variety of paths  from the edge state
leading to the turbulence. Besides demonstrating the utility of the 
symmetry reduction in both $z$ and $\theta$, this computation
also shows how rich the dynamics on the laminar-turbulent boundary
can be. One main message we would like to deliver with the aid of 
these illustrations is that the asymptotic dynamics on the 
laminar-turbulent boundary in pipe flow should not be treated as a
single state;	 but the edge state contains different regions with 
qualitatively different dynamics dictated by the nearby invariant 
solutions.

Both traveling waves we studied in this paper belonged to the 
shift-and-reflect invariant subspace of the pipe flow. One might 
argue against their relevance for the dynamics in the 
laminar-turbulent boundary since generic trajectories live in the 
full \statesp. Cvitanovi\'c \etal \rf{SCD07} nicely illustrates
for one-dimensional Kuramoto-Sivashinsky system that the unstable 
manifolds of equilibrium solutions, all of which belong to the 
reflection invariant subspace, successfully captures the qualitative
dynamics in the full \statesp\ of the \On{2}-equivariant system. Our 
case here is similar to theirs since the symmetry in the azimuthal 
direction is also \On{2}. Therefore, it is quite reasonable 
to expect for \Sone's unstable manifold to form the backbone of the 
asymptotic dynamics on the laminar-turbulent boundary, given all the 
evidence that the edge state is located in its vicinity. 
Indeed, visualizations of \reffig{f-minDistLower} clearly show 
the similarities between the full \statesp\ trajectories and 
the traveling waves nearby.

Recently, Suri \etal \rf{STGS16} studied the weakly turbulent 
quasi-two-dimensional flow experimentally and numerically. 
They demonstrated that when a turbulent trajectory comes close to an 
equilibrium, it leaves this neighborhood by following the unstable manifold 
of the solution. The tools we presented here, in particular, the methods for
approximating and visualizing the unstable manifolds after 
symmetry-reduction, pave the way for a similar analysis in 
three-dimensional pressure-driven flows, which have traveling wave 
solutions rather than equilibria. 

As we have shown in 
\refsect{s-numexp}, symmetry-reduction frees us from 
reduced metrics, such as energy input or dissipation, 
that do not carry all information of the \statesp, 
and allows for measuring distances in the full \statesp . 
However, one should always keep in 
mind that two points that are at a short distance in the \statesp\ 
of a nonlinear system might evolve towards completely different 
regions. In order to conclusively answer whether or not a \statesp\ 
trajectory is ``guided'' by a particular unstable manifold, we should
measure distances between curves in the \statesp, rather
than points. While the finite-dimensional projections such as 
figures \ref{f-man1winst}, \ref{f-man3DS1}, and \ref{f-s1nManifolds}
serve this purpose, they do not provide a complete picture of the 
infinite-dimensional \statesp . We believe developing computationally
feasible methods for comparing shapes in high-dimensional \statesp s 
is an important future problem for turbulence studies.

\acknowledgements

We would like to acknowledge stimulating discussions with 
Yohann Duguet and Predrag Cvitanovi\'c.
This research was supported in part by the National Science
Foundation under Grant No. NSF PHY11-25915.


\bibliography{../../bibtex/neubauten}

\begin{thebibliography}{36}%
\makeatletter
\providecommand \@ifxundefined [1]{%
 \@ifx{#1\undefined}
}%
\providecommand \@ifnum [1]{%
 \ifnum #1\expandafter \@firstoftwo
 \else \expandafter \@secondoftwo
 \fi
}%
\providecommand \@ifx [1]{%
 \ifx #1\expandafter \@firstoftwo
 \else \expandafter \@secondoftwo
 \fi
}%
\providecommand \natexlab [1]{#1}%
\providecommand \enquote  [1]{``#1''}%
\providecommand \bibnamefont  [1]{#1}%
\providecommand \bibfnamefont [1]{#1}%
\providecommand \citenamefont [1]{#1}%
\providecommand \href@noop [0]{\@secondoftwo}%
\providecommand \href [0]{\begingroup \@sanitize@url \@href}%
\providecommand \@href[1]{\@@startlink{#1}\@@href}%
\providecommand \@@href[1]{\endgroup#1\@@endlink}%
\providecommand \@sanitize@url [0]{\catcode `\\12\catcode `\$12\catcode
  `\&12\catcode `\#12\catcode `\^12\catcode `\_12\catcode `\%12\relax}%
\providecommand \@@startlink[1]{}%
\providecommand \@@endlink[0]{}%
\providecommand \url  [0]{\begingroup\@sanitize@url \@url }%
\providecommand \@url [1]{\endgroup\@href {#1}{\urlprefix }}%
\providecommand \urlprefix  [0]{URL }%
\providecommand \Eprint [0]{\href }%
\providecommand \doibase [0]{http://dx.doi.org/}%
\providecommand \selectlanguage [0]{\@gobble}%
\providecommand \bibinfo  [0]{\@secondoftwo}%
\providecommand \bibfield  [0]{\@secondoftwo}%
\providecommand \translation [1]{[#1]}%
\providecommand \BibitemOpen [0]{}%
\providecommand \bibitemStop [0]{}%
\providecommand \bibitemNoStop [0]{.\EOS\space}%
\providecommand \EOS [0]{\spacefactor3000\relax}%
\providecommand \BibitemShut  [1]{\csname bibitem#1\endcsname}%
\let\auto@bib@innerbib\@empty
\bibitem [{\citenamefont {Meseguer}\ and\ \citenamefont
  {Trefethen}(2003)}]{MesTre03}%
  \BibitemOpen
  \bibfield  {author} {\bibinfo {author} {\bibfnamefont {A}~\bibnamefont
  {Meseguer}}\ and\ \bibinfo {author} {\bibfnamefont {L.~N}\ \bibnamefont
  {Trefethen}},\ }\bibfield  {title} {\enquote {\bibinfo {title} {Linearized
  pipe flow to {Reynolds} number $10^7$},}\ }\href@noop {} {\bibfield
  {journal} {\bibinfo  {journal} {J. Comput. Phys.}\ }\textbf {\bibinfo
  {volume} {186}},\ \bibinfo {pages} {178--197} (\bibinfo {year}
  {2003})}\BibitemShut {NoStop}%
\bibitem [{\citenamefont {Faisst}\ and\ \citenamefont {Eckhardt}(2003)}]{FE03}%
  \BibitemOpen
  \bibfield  {author} {\bibinfo {author} {\bibfnamefont {H.}~\bibnamefont
  {Faisst}}\ and\ \bibinfo {author} {\bibfnamefont {B.}~\bibnamefont
  {Eckhardt}},\ }\bibfield  {title} {\enquote {\bibinfo {title} {Traveling
  waves in pipe flow},}\ }\href@noop {} {\bibfield  {journal} {\bibinfo
  {journal} {Phys. Rev. Lett.}\ }\textbf {\bibinfo {volume} {91}},\ \bibinfo
  {pages} {224502} (\bibinfo {year} {2003})}\BibitemShut {NoStop}%
\bibitem [{\citenamefont {Pringle}\ and\ \citenamefont
  {Kerswell}(2007)}]{Pringle07}%
  \BibitemOpen
  \bibfield  {author} {\bibinfo {author} {\bibfnamefont {C.~C.~T.}\
  \bibnamefont {Pringle}}\ and\ \bibinfo {author} {\bibfnamefont {R.~R.}\
  \bibnamefont {Kerswell}},\ }\bibfield  {title} {\enquote {\bibinfo {title}
  {Asymmetric, helical, and mirror-symmetric traveling waves in pipe flow},}\
  }\href@noop {} {\bibfield  {journal} {\bibinfo  {journal} {Phys. Rev. Lett.}\
  }\textbf {\bibinfo {volume} {99}},\ \bibinfo {pages} {074502} (\bibinfo
  {year} {2007})}\BibitemShut {NoStop}%
\bibitem [{\citenamefont {Pringle}\ \emph {et~al.}(2009)\citenamefont
  {Pringle}, \citenamefont {Duguet},\ and\ \citenamefont
  {Kerswell}}]{Pringle09}%
  \BibitemOpen
  \bibfield  {author} {\bibinfo {author} {\bibfnamefont {C.~C.~T.}\
  \bibnamefont {Pringle}}, \bibinfo {author} {\bibfnamefont {Y.}~\bibnamefont
  {Duguet}}, \ and\ \bibinfo {author} {\bibfnamefont {R.~R.}\ \bibnamefont
  {Kerswell}},\ }\bibfield  {title} {\enquote {\bibinfo {title} {Highly
  symmetric travelling waves in pipe flow},}\ }\href@noop {} {\bibfield
  {journal} {\bibinfo  {journal} {Phil. Trans. Royal Soc. A}\ }\textbf
  {\bibinfo {volume} {367}},\ \bibinfo {pages} {457--472} (\bibinfo {year}
  {2009})},\ \bibinfo {note} {\arXiv{0804.4854}}\BibitemShut {NoStop}%
\bibitem [{\citenamefont {Schneider}\ \emph {et~al.}(2007)\citenamefont
  {Schneider}, \citenamefont {Eckhardt},\ and\ \citenamefont
  {Yorke}}]{SchEckYor07}%
  \BibitemOpen
  \bibfield  {author} {\bibinfo {author} {\bibfnamefont {T.~M.}\ \bibnamefont
  {Schneider}}, \bibinfo {author} {\bibfnamefont {B.}~\bibnamefont {Eckhardt}},
  \ and\ \bibinfo {author} {\bibfnamefont {J.}~\bibnamefont {Yorke}},\
  }\bibfield  {title} {\enquote {\bibinfo {title} {Turbulence, transition, and
  the edge of chaos in pipe flow},}\ }\href@noop {} {\bibfield  {journal}
  {\bibinfo  {journal} {Phys. Rev. Lett.}\ }\textbf {\bibinfo {volume} {99}},\
  \bibinfo {pages} {034502} (\bibinfo {year} {2007})}\BibitemShut {NoStop}%
\bibitem [{\citenamefont {Duguet}\ \emph {et~al.}(2008)\citenamefont {Duguet},
  \citenamefont {Willis},\ and\ \citenamefont {Kerswell}}]{duguet07}%
  \BibitemOpen
  \bibfield  {author} {\bibinfo {author} {\bibfnamefont {Y.}~\bibnamefont
  {Duguet}}, \bibinfo {author} {\bibfnamefont {A.~P.}\ \bibnamefont {Willis}},
  \ and\ \bibinfo {author} {\bibfnamefont {R.~R.}\ \bibnamefont {Kerswell}},\
  }\bibfield  {title} {\enquote {\bibinfo {title} {Transition in pipe flow: the
  saddle structure on the boundary of turbulence},}\ }\href {\doibase
  10.1017/S0022112008003248} {\bibfield  {journal} {\bibinfo  {journal} {J.
  Fluid Mech.}\ }\textbf {\bibinfo {volume} {613}},\ \bibinfo {pages}
  {255--274} (\bibinfo {year} {2008})},\ \bibinfo {note}
  {\arXiv{0711.2175}}\BibitemShut {NoStop}%
\bibitem [{\citenamefont {Mellibovsky}\ \emph {et~al.}(2009)\citenamefont
  {Mellibovsky}, \citenamefont {Meseguer}, \citenamefont {Schneider},\ and\
  \citenamefont {Eckhardt}}]{MMSE09}%
  \BibitemOpen
  \bibfield  {author} {\bibinfo {author} {\bibfnamefont {F.}~\bibnamefont
  {Mellibovsky}}, \bibinfo {author} {\bibfnamefont {A.}~\bibnamefont
  {Meseguer}}, \bibinfo {author} {\bibfnamefont {T.~M.}\ \bibnamefont
  {Schneider}}, \ and\ \bibinfo {author} {\bibfnamefont {B.}~\bibnamefont
  {Eckhardt}},\ }\bibfield  {title} {\enquote {\bibinfo {title} {Transition in
  localized pipe flow turbulence},}\ }\href {\doibase
  10.1103/PhysRevLett.103.054502} {\bibfield  {journal} {\bibinfo  {journal}
  {Phys. Rev. Lett.}\ }\textbf {\bibinfo {volume} {103}},\ \bibinfo {pages}
  {054502} (\bibinfo {year} {2009})}\BibitemShut {NoStop}%
\bibitem [{\citenamefont {Hopf}(1948)}]{hopf48}%
  \BibitemOpen
  \bibfield  {author} {\bibinfo {author} {\bibfnamefont {E.}~\bibnamefont
  {Hopf}},\ }\bibfield  {title} {\enquote {\bibinfo {title} {A mathematical
  example displaying features of turbulence},}\ }\href {\doibase
  10.1002/cpa.3160010401} {\bibfield  {journal} {\bibinfo  {journal} {Commun.
  Pure Appl. Math.}\ }\textbf {\bibinfo {volume} {1}},\ \bibinfo {pages}
  {303--322} (\bibinfo {year} {1948})}\BibitemShut {NoStop}%
\bibitem [{\citenamefont {Itano}\ and\ \citenamefont {Toh}(2001)}]{IT01}%
  \BibitemOpen
  \bibfield  {author} {\bibinfo {author} {\bibfnamefont {T.}~\bibnamefont
  {Itano}}\ and\ \bibinfo {author} {\bibfnamefont {S.}~\bibnamefont {Toh}},\
  }\bibfield  {title} {\enquote {\bibinfo {title} {The dynamics of bursting
  process in wall turbulence},}\ }\href@noop {} {\bibfield  {journal} {\bibinfo
   {journal} {J. Phys. Soc. Japan}\ }\textbf {\bibinfo {volume} {70}},\
  \bibinfo {pages} {701--714} (\bibinfo {year} {2001})}\BibitemShut {NoStop}%
\bibitem [{\citenamefont {Toh}\ and\ \citenamefont {Itano}(2003)}]{TI03}%
  \BibitemOpen
  \bibfield  {author} {\bibinfo {author} {\bibfnamefont {S.}~\bibnamefont
  {Toh}}\ and\ \bibinfo {author} {\bibfnamefont {T.}~\bibnamefont {Itano}},\
  }\bibfield  {title} {\enquote {\bibinfo {title} {A periodic-like solution in
  channel flow},}\ }\href@noop {} {\bibfield  {journal} {\bibinfo  {journal}
  {J. Fluid Mech.}\ }\textbf {\bibinfo {volume} {481}},\ \bibinfo {pages}
  {67--76} (\bibinfo {year} {2003})}\BibitemShut {NoStop}%
\bibitem [{\citenamefont {Schneider}\ \emph {et~al.}(2008)\citenamefont
  {Schneider}, \citenamefont {Gibson}, \citenamefont {Lagha}, \citenamefont
  {Lillo},\ and\ \citenamefont {Eckhardt}}]{SGLDE08}%
  \BibitemOpen
  \bibfield  {author} {\bibinfo {author} {\bibfnamefont {T.~M.}\ \bibnamefont
  {Schneider}}, \bibinfo {author} {\bibfnamefont {J.~F.}\ \bibnamefont
  {Gibson}}, \bibinfo {author} {\bibfnamefont {M.}~\bibnamefont {Lagha}},
  \bibinfo {author} {\bibfnamefont {F.~De}\ \bibnamefont {Lillo}}, \ and\
  \bibinfo {author} {\bibfnamefont {B.}~\bibnamefont {Eckhardt}},\ }\bibfield
  {title} {\enquote {\bibinfo {title} {Laminar-turbulent boundary in plane
  {Couette} flow},}\ }\href@noop {} {\bibfield  {journal} {\bibinfo  {journal}
  {Phys. Rev. E.}\ }\textbf {\bibinfo {volume} {78}},\ \bibinfo {pages}
  {037301} (\bibinfo {year} {2008})},\ \bibinfo {note}
  {\arXiv{0805.1015}}\BibitemShut {NoStop}%
\bibitem [{\citenamefont {Schneider}\ \emph {et~al.}(2010)\citenamefont
  {Schneider}, \citenamefont {Marinc},\ and\ \citenamefont
  {Eckhardt}}]{ScMaEc10}%
  \BibitemOpen
  \bibfield  {author} {\bibinfo {author} {\bibfnamefont {T.~M.}\ \bibnamefont
  {Schneider}}, \bibinfo {author} {\bibfnamefont {D.}~\bibnamefont {Marinc}}, \
  and\ \bibinfo {author} {\bibfnamefont {B.}~\bibnamefont {Eckhardt}},\
  }\bibfield  {title} {\enquote {\bibinfo {title} {Localized edge states
  nucleate turbulence in extended plane {Couette} cells},}\ }\href {\doibase
  10.1017/S0022112009993144} {\bibfield  {journal} {\bibinfo  {journal} {J.
  Fluid Mech.}\ }\textbf {\bibinfo {volume} {646}},\ \bibinfo {pages}
  {441--451} (\bibinfo {year} {2010})}\BibitemShut {NoStop}%
\bibitem [{\citenamefont {Zammert}\ and\ \citenamefont
  {Eckhardt}(2014)}]{ZamEck14a}%
  \BibitemOpen
  \bibfield  {author} {\bibinfo {author} {\bibfnamefont {S.}~\bibnamefont
  {Zammert}}\ and\ \bibinfo {author} {\bibfnamefont {B.}~\bibnamefont
  {Eckhardt}},\ }\bibfield  {title} {\enquote {\bibinfo {title} {A spotlike
  edge state in plane {Poiseuille} flow},}\ }\href {\doibase
  10.1002/pamm.201410283} {\bibfield  {journal} {\bibinfo  {journal} {PAMM}\
  }\textbf {\bibinfo {volume} {14}},\ \bibinfo {pages} {591--592} (\bibinfo
  {year} {2014})}\BibitemShut {NoStop}%
\bibitem [{\citenamefont {Khapko}\ \emph {et~al.}(2016)\citenamefont {Khapko},
  \citenamefont {Kreilos}, \citenamefont {Schlatter}, \citenamefont {Duguet},
  \citenamefont {Eckhardt},\ and\ \citenamefont {Henningson}}]{KKSDEH2016}%
  \BibitemOpen
  \bibfield  {author} {\bibinfo {author} {\bibfnamefont {T.}~\bibnamefont
  {Khapko}}, \bibinfo {author} {\bibfnamefont {T.}~\bibnamefont {Kreilos}},
  \bibinfo {author} {\bibfnamefont {P.}~\bibnamefont {Schlatter}}, \bibinfo
  {author} {\bibfnamefont {Y.}~\bibnamefont {Duguet}}, \bibinfo {author}
  {\bibfnamefont {B.}~\bibnamefont {Eckhardt}}, \ and\ \bibinfo {author}
  {\bibfnamefont {D.~S.}\ \bibnamefont {Henningson}},\ }\bibfield  {title}
  {\enquote {\bibinfo {title} {Edge states as mediators of bypass transition in
  boundary-layer flows},}\ }\href {\doibase 10.1017/jfm.2016.434} {\bibfield
  {journal} {\bibinfo  {journal} {J. Fluid Mech.}\ }\textbf {\bibinfo {volume}
  {801}} (\bibinfo {year} {2016}),\ 10.1017/jfm.2016.434}\BibitemShut {NoStop}%
\bibitem [{\citenamefont {Cvitanovi{\'c}}\ \emph {et~al.}(2015)\citenamefont
  {Cvitanovi{\'c}}, \citenamefont {Artuso}, \citenamefont {Mainieri},
  \citenamefont {Tanner},\ and\ \citenamefont {Vattay}}]{DasBuch}%
  \BibitemOpen
  \bibfield  {author} {\bibinfo {author} {\bibfnamefont {P.}~\bibnamefont
  {Cvitanovi{\'c}}}, \bibinfo {author} {\bibfnamefont {R.}~\bibnamefont
  {Artuso}}, \bibinfo {author} {\bibfnamefont {R.}~\bibnamefont {Mainieri}},
  \bibinfo {author} {\bibfnamefont {G.}~\bibnamefont {Tanner}}, \ and\ \bibinfo
  {author} {\bibfnamefont {G.}~\bibnamefont {Vattay}},\ }\href@noop {} {\emph
  {\bibinfo {title} {Chaos: Classical and Quantum}}}\ (\bibinfo  {publisher}
  {Niels Bohr Inst.},\ \bibinfo {address} {Copenhagen},\ \bibinfo {year}
  {2015})\ \bibinfo {note} {{\tt ChaosBook.org}}\BibitemShut {NoStop}%
\bibitem [{\citenamefont {Kawahara}\ and\ \citenamefont
  {Kida}(2001)}]{KawKida01}%
  \BibitemOpen
  \bibfield  {author} {\bibinfo {author} {\bibfnamefont {G.}~\bibnamefont
  {Kawahara}}\ and\ \bibinfo {author} {\bibfnamefont {S.}~\bibnamefont
  {Kida}},\ }\bibfield  {title} {\enquote {\bibinfo {title} {Periodic motion
  embedded in plane {Couette} turbulence: {Regeneration} cycle and burst},}\
  }\href@noop {} {\bibfield  {journal} {\bibinfo  {journal} {J. Fluid Mech.}\
  }\textbf {\bibinfo {volume} {449}},\ \bibinfo {pages} {291--300} (\bibinfo
  {year} {2001})}\BibitemShut {NoStop}%
\bibitem [{\citenamefont {Viswanath}(2007)}]{Visw07b}%
  \BibitemOpen
  \bibfield  {author} {\bibinfo {author} {\bibfnamefont {D.}~\bibnamefont
  {Viswanath}},\ }\bibfield  {title} {\enquote {\bibinfo {title} {Recurrent
  motions within plane {Couette} turbulence},}\ }\href {\doibase
  10.1017/S0022112007005459} {\bibfield  {journal} {\bibinfo  {journal} {J.
  Fluid Mech.}\ }\textbf {\bibinfo {volume} {580}},\ \bibinfo {pages}
  {339--358} (\bibinfo {year} {2007})},\ \bibinfo {note}
  {\arXiv{physics/0604062}}\BibitemShut {NoStop}%
\bibitem [{\citenamefont {Ritter}\ \emph {et~al.}(2016)\citenamefont {Ritter},
  \citenamefont {Mellibovsky},\ and\ \citenamefont {Avila}}]{RiMeAv16}%
  \BibitemOpen
  \bibfield  {author} {\bibinfo {author} {\bibfnamefont {P.}~\bibnamefont
  {Ritter}}, \bibinfo {author} {\bibfnamefont {F.}~\bibnamefont {Mellibovsky}},
  \ and\ \bibinfo {author} {\bibfnamefont {M.}~\bibnamefont {Avila}},\
  }\bibfield  {title} {\enquote {\bibinfo {title} {Emergence of spatio-temporal
  dynamics from exact coherent solutions in pipe flow},}\ }\href {\doibase
  10.1088/1367-2630/18/8/083031} {\bibfield  {journal} {\bibinfo  {journal}
  {New J. Phys.}\ }\textbf {\bibinfo {volume} {18}},\ \bibinfo {pages} {083031}
  (\bibinfo {year} {2016})}\BibitemShut {NoStop}%
\bibitem [{\citenamefont {Gibson}\ \emph {et~al.}(2008)\citenamefont {Gibson},
  \citenamefont {Halcrow},\ and\ \citenamefont {Cvitanovi{\'c}}}]{GHCW07}%
  \BibitemOpen
  \bibfield  {author} {\bibinfo {author} {\bibfnamefont {J.~F.}\ \bibnamefont
  {Gibson}}, \bibinfo {author} {\bibfnamefont {J.}~\bibnamefont {Halcrow}}, \
  and\ \bibinfo {author} {\bibfnamefont {P.}~\bibnamefont {Cvitanovi{\'c}}},\
  }\bibfield  {title} {\enquote {\bibinfo {title} {Visualizing the geometry of
  state space in plane {Couette} flow},}\ }\href {\doibase
  10.1017/S002211200800267X} {\bibfield  {journal} {\bibinfo  {journal} {J.
  Fluid Mech.}\ }\textbf {\bibinfo {volume} {611}},\ \bibinfo {pages}
  {107--130} (\bibinfo {year} {2008})},\ \bibinfo {note}
  {\arXiv{0705.3957}}\BibitemShut {NoStop}%
\bibitem [{\citenamefont {Budanur}\ and\ \citenamefont {Hof}(2017)}]{BudHof17}%
  \BibitemOpen
  \bibfield  {author} {\bibinfo {author} {\bibfnamefont {N.~B.}\ \bibnamefont
  {Budanur}}\ and\ \bibinfo {author} {\bibfnamefont {B.}~\bibnamefont {Hof}},\
  }\bibfield  {title} {\enquote {\bibinfo {title} {Heteroclinic path to
  spatially localized chaos in pipe flow},}\ }\href {\doibase
  10.1017/jfm.2017.516} {\bibfield  {journal} {\bibinfo  {journal} {J. Fluid
  Mech}\ }\textbf {\bibinfo {volume} {827, R1}} (\bibinfo {year} {2017}),\
  10.1017/jfm.2017.516}\BibitemShut {NoStop}%
\bibitem [{\citenamefont {Willis}(2017)}]{Willis2017}%
  \BibitemOpen
  \bibfield  {author} {\bibinfo {author} {\bibfnamefont {A.~P.}\ \bibnamefont
  {Willis}},\ }\bibfield  {title} {\enquote {\bibinfo {title} {The
  {Openpipeflow} {Navier-Stokes} solver},}\ }\href {\doibase
  https://doi.org/10.1016/j.softx.2017.05.003} {\bibfield  {journal} {\bibinfo
  {journal} {SoftwareX}\ }\textbf {\bibinfo {volume} {6}},\ \bibinfo {pages}
  {124 -- 127} (\bibinfo {year} {2017})}\BibitemShut {NoStop}%
\bibitem [{\citenamefont {Wedin}\ and\ \citenamefont {Kerswell}(2004)}]{WK04}%
  \BibitemOpen
  \bibfield  {author} {\bibinfo {author} {\bibfnamefont {H.}~\bibnamefont
  {Wedin}}\ and\ \bibinfo {author} {\bibfnamefont {R.~R.}\ \bibnamefont
  {Kerswell}},\ }\bibfield  {title} {\enquote {\bibinfo {title} {Exact coherent
  structures in pipe flow: {Traveling} wave solutions},}\ }\href@noop {}
  {\bibfield  {journal} {\bibinfo  {journal} {J. Fluid Mech.}\ }\textbf
  {\bibinfo {volume} {508}},\ \bibinfo {pages} {333--371} (\bibinfo {year}
  {2004})}\BibitemShut {NoStop}%
\bibitem [{\citenamefont {Budanur}\ \emph {et~al.}(2017)\citenamefont
  {Budanur}, \citenamefont {Short}, \citenamefont {Farazmand}, \citenamefont
  {Willis},\ and\ \citenamefont {Cvitanovi{\'c}}}]{WFSBC15}%
  \BibitemOpen
  \bibfield  {author} {\bibinfo {author} {\bibfnamefont {N.~B.}\ \bibnamefont
  {Budanur}}, \bibinfo {author} {\bibfnamefont {K.~Y.}\ \bibnamefont {Short}},
  \bibinfo {author} {\bibfnamefont {M.}~\bibnamefont {Farazmand}}, \bibinfo
  {author} {\bibfnamefont {A.~P.}\ \bibnamefont {Willis}}, \ and\ \bibinfo
  {author} {\bibfnamefont {P.}~\bibnamefont {Cvitanovi{\'c}}},\ }\bibfield
  {title} {\enquote {\bibinfo {title} {Relative periodic orbits form the
  backbone of turbulent pipe flow},}\ }\href {\doibase 10.1017/jfm.2017.699}
  {\bibfield  {journal} {\bibinfo  {journal} {J. Fluid Mech}\ }\textbf
  {\bibinfo {volume} {833}} (\bibinfo {year} {2017}),\
  10.1017/jfm.2017.699}\BibitemShut {NoStop}%
\bibitem [{\citenamefont {Willis}\ \emph {et~al.}(2013)\citenamefont {Willis},
  \citenamefont {Cvitanovi{\'c}},\ and\ \citenamefont {Avila}}]{ACHKW11}%
  \BibitemOpen
  \bibfield  {author} {\bibinfo {author} {\bibfnamefont {A.~P.}\ \bibnamefont
  {Willis}}, \bibinfo {author} {\bibfnamefont {P.}~\bibnamefont
  {Cvitanovi{\'c}}}, \ and\ \bibinfo {author} {\bibfnamefont {M.}~\bibnamefont
  {Avila}},\ }\bibfield  {title} {\enquote {\bibinfo {title} {Revealing the
  state space of turbulent pipe flow by symmetry reduction},}\ }\href {\doibase
  10.1017/jfm.2013.75} {\bibfield  {journal} {\bibinfo  {journal} {J. Fluid
  Mech.}\ }\textbf {\bibinfo {volume} {721}},\ \bibinfo {pages} {514--540}
  (\bibinfo {year} {2013})},\ \bibinfo {note} {\arXiv{1203.3701}}\BibitemShut
  {NoStop}%
\bibitem [{\citenamefont {Budanur}\ \emph
  {et~al.}(2015{\natexlab{a}})\citenamefont {Budanur}, \citenamefont
  {Cvitanovi\'c}, \citenamefont {Davidchack},\ and\ \citenamefont
  {Siminos}}]{BudCvi14}%
  \BibitemOpen
  \bibfield  {author} {\bibinfo {author} {\bibfnamefont {N.~B.}\ \bibnamefont
  {Budanur}}, \bibinfo {author} {\bibfnamefont {P.}~\bibnamefont
  {Cvitanovi\'c}}, \bibinfo {author} {\bibfnamefont {R.~L.}\ \bibnamefont
  {Davidchack}}, \ and\ \bibinfo {author} {\bibfnamefont {E.}~\bibnamefont
  {Siminos}},\ }\bibfield  {title} {\enquote {\bibinfo {title} {Reduction of
  the {SO(2)} symmetry for spatially extended dynamical systems},}\ }\href
  {\doibase 10.1103/PhysRevLett.114.084102} {\bibfield  {journal} {\bibinfo
  {journal} {Phys. Rev. Lett.}\ }\textbf {\bibinfo {volume} {114}},\ \bibinfo
  {pages} {084102} (\bibinfo {year} {2015}{\natexlab{a}})}\BibitemShut
  {NoStop}%
\bibitem [{\citenamefont {Budanur}\ \emph
  {et~al.}(2015{\natexlab{b}})\citenamefont {Budanur}, \citenamefont
  {Borrero-Echeverry},\ and\ \citenamefont {Cvitanovi\'c}}]{BuBoCvSi14}%
  \BibitemOpen
  \bibfield  {author} {\bibinfo {author} {\bibfnamefont {N.~B.}\ \bibnamefont
  {Budanur}}, \bibinfo {author} {\bibfnamefont {D.}~\bibnamefont
  {Borrero-Echeverry}}, \ and\ \bibinfo {author} {\bibfnamefont
  {P.}~\bibnamefont {Cvitanovi\'c}},\ }\bibfield  {title} {\enquote {\bibinfo
  {title} {Periodic orbit analysis of a system with continuous symmetry - a
  tutorial},}\ }\href {\doibase 10.1063/1.4923742} {\bibfield  {journal}
  {\bibinfo  {journal} {Chaos}\ }\textbf {\bibinfo {volume} {25}},\ \bibinfo
  {pages} {073112} (\bibinfo {year} {2015}{\natexlab{b}})},\ \bibinfo {note}
  {\arXiv{1411.3303}}\BibitemShut {NoStop}%
\bibitem [{\citenamefont {Froehlich}\ and\ \citenamefont
  {Cvitanovi{\'c}}(2012)}]{FrCv11}%
  \BibitemOpen
  \bibfield  {author} {\bibinfo {author} {\bibfnamefont {S.}~\bibnamefont
  {Froehlich}}\ and\ \bibinfo {author} {\bibfnamefont {P.}~\bibnamefont
  {Cvitanovi{\'c}}},\ }\bibfield  {title} {\enquote {\bibinfo {title}
  {Reduction of continuous symmetries of chaotic flows by the method of
  slices},}\ }\href {\doibase 10.1016/j.cnsns.2011.07.007} {\bibfield
  {journal} {\bibinfo  {journal} {Commun. Nonlinear Sci. Numer. Simul.}\
  }\textbf {\bibinfo {volume} {17}},\ \bibinfo {pages} {2074--2084} (\bibinfo
  {year} {2012})},\ \bibinfo {note} {\arXiv{1101.3037}}\BibitemShut {NoStop}%
\bibitem [{\citenamefont {Cvitanovi\'c}\ \emph {et~al.}(2012)\citenamefont
  {Cvitanovi\'c}, \citenamefont {Borrero-Echeverry}, \citenamefont {Carroll},
  \citenamefont {Robbins},\ and\ \citenamefont {Siminos}}]{atlas12}%
  \BibitemOpen
  \bibfield  {author} {\bibinfo {author} {\bibfnamefont {P.}~\bibnamefont
  {Cvitanovi\'c}}, \bibinfo {author} {\bibfnamefont {D.}~\bibnamefont
  {Borrero-Echeverry}}, \bibinfo {author} {\bibfnamefont {K.}~\bibnamefont
  {Carroll}}, \bibinfo {author} {\bibfnamefont {B.}~\bibnamefont {Robbins}}, \
  and\ \bibinfo {author} {\bibfnamefont {E.}~\bibnamefont {Siminos}},\
  }\bibfield  {title} {\enquote {\bibinfo {title} {Cartography of
  high-dimensional flows: {A} visual guide to sections and slices},}\ }\href
  {\doibase 10.1063/1.4758309} {\bibfield  {journal} {\bibinfo  {journal}
  {Chaos}\ }\textbf {\bibinfo {volume} {22}},\ \bibinfo {pages} {047506}
  (\bibinfo {year} {2012})}\BibitemShut {NoStop}%
\bibitem [{\citenamefont {Rowley}\ and\ \citenamefont
  {Marsden}(2000)}]{rowley_reconstruction_2000}%
  \BibitemOpen
  \bibfield  {author} {\bibinfo {author} {\bibfnamefont {C.~W.}\ \bibnamefont
  {Rowley}}\ and\ \bibinfo {author} {\bibfnamefont {J.~E.}\ \bibnamefont
  {Marsden}},\ }\bibfield  {title} {\enquote {\bibinfo {title} {Reconstruction
  equations and the {Karhunen-Lo{\'e}ve} expansion for systems with
  symmetry},}\ }\href {\doibase 10.1016/S0167-2789(00)00042-7} {\bibfield
  {journal} {\bibinfo  {journal} {Physica D}\ }\textbf {\bibinfo {volume}
  {142}},\ \bibinfo {pages} {1--19} (\bibinfo {year} {2000})}\BibitemShut
  {NoStop}%
\bibitem [{\citenamefont {Chossat}\ and\ \citenamefont
  {Lauterbach}(2000)}]{ChossLaut00}%
  \BibitemOpen
  \bibfield  {author} {\bibinfo {author} {\bibfnamefont {P.}~\bibnamefont
  {Chossat}}\ and\ \bibinfo {author} {\bibfnamefont {R.}~\bibnamefont
  {Lauterbach}},\ }\href@noop {} {\emph {\bibinfo {title} {Methods in
  Equivariant Bifurcations and Dynamical Systems}}}\ (\bibinfo  {publisher}
  {World Scientific},\ \bibinfo {address} {Singapore},\ \bibinfo {year}
  {2000})\BibitemShut {NoStop}%
\bibitem [{\citenamefont {Armbruster}\ \emph {et~al.}(1988)\citenamefont
  {Armbruster}, \citenamefont {Guckenheimer},\ and\ \citenamefont
  {Holmes}}]{AGHks88}%
  \BibitemOpen
  \bibfield  {author} {\bibinfo {author} {\bibfnamefont {D.}~\bibnamefont
  {Armbruster}}, \bibinfo {author} {\bibfnamefont {J.}~\bibnamefont
  {Guckenheimer}}, \ and\ \bibinfo {author} {\bibfnamefont {P.}~\bibnamefont
  {Holmes}},\ }\bibfield  {title} {\enquote {\bibinfo {title} {Heteroclinic
  cycles and modulated travelling waves in systems with {O(2)} symmetry},}\
  }\href@noop {} {\bibfield  {journal} {\bibinfo  {journal} {Physica D}\
  }\textbf {\bibinfo {volume} {29}},\ \bibinfo {pages} {257--282} (\bibinfo
  {year} {1988})}\BibitemShut {NoStop}%
\bibitem [{\citenamefont {de~Lozar}\ \emph {et~al.}(2012)\citenamefont
  {de~Lozar}, \citenamefont {Mellibovsky}, \citenamefont {Avila},\ and\
  \citenamefont {Hof}}]{deLMeAvHo12}%
  \BibitemOpen
  \bibfield  {author} {\bibinfo {author} {\bibfnamefont {A.}~\bibnamefont
  {de~Lozar}}, \bibinfo {author} {\bibfnamefont {F.}~\bibnamefont
  {Mellibovsky}}, \bibinfo {author} {\bibfnamefont {M.}~\bibnamefont {Avila}},
  \ and\ \bibinfo {author} {\bibfnamefont {B.}~\bibnamefont {Hof}},\ }\bibfield
   {title} {\enquote {\bibinfo {title} {Edge state in pipe flow experiments},}\
  }\href {\doibase 10.1103/PhysRevLett.108.214502} {\bibfield  {journal}
  {\bibinfo  {journal} {Phys. Rev. Lett.}\ }\textbf {\bibinfo {volume} {108}},\
  \bibinfo {pages} {214502} (\bibinfo {year} {2012})}\BibitemShut {NoStop}%
\bibitem [{\citenamefont {Krauskopf}\ \emph {et~al.}(2005)\citenamefont
  {Krauskopf}, \citenamefont {Osinga}, \citenamefont {Doedel}, \citenamefont
  {Henderson}, \citenamefont {Guckenheimer}, \citenamefont {Vladimirsky},
  \citenamefont {Dellnitz},\ and\ \citenamefont {Junge}}]{Krauskopf2005}%
  \BibitemOpen
  \bibfield  {author} {\bibinfo {author} {\bibfnamefont {B.}~\bibnamefont
  {Krauskopf}}, \bibinfo {author} {\bibfnamefont {H.~M.}\ \bibnamefont
  {Osinga}}, \bibinfo {author} {\bibfnamefont {E.~J.}\ \bibnamefont {Doedel}},
  \bibinfo {author} {\bibfnamefont {M.~E.}\ \bibnamefont {Henderson}}, \bibinfo
  {author} {\bibfnamefont {J.}~\bibnamefont {Guckenheimer}}, \bibinfo {author}
  {\bibfnamefont {A.}~\bibnamefont {Vladimirsky}}, \bibinfo {author}
  {\bibfnamefont {M.}~\bibnamefont {Dellnitz}}, \ and\ \bibinfo {author}
  {\bibfnamefont {O.}~\bibnamefont {Junge}},\ }\bibfield  {title} {\enquote
  {\bibinfo {title} {A survey of methods for computing (un)stable manifolds of
  vector fields},}\ }\href {\doibase 10.1142/S0218127405012533} {\bibfield
  {journal} {\bibinfo  {journal} {Int. J. Bifurc. Chaos}\ }\textbf {\bibinfo
  {volume} {15}},\ \bibinfo {pages} {763--791} (\bibinfo {year}
  {2005})}\BibitemShut {NoStop}%
\bibitem [{\citenamefont {Skufca}\ \emph {et~al.}(2006)\citenamefont {Skufca},
  \citenamefont {Yorke},\ and\ \citenamefont {Eckhardt}}]{SYE05}%
  \BibitemOpen
  \bibfield  {author} {\bibinfo {author} {\bibfnamefont {J.~D.}\ \bibnamefont
  {Skufca}}, \bibinfo {author} {\bibfnamefont {J.~A.}\ \bibnamefont {Yorke}}, \
  and\ \bibinfo {author} {\bibfnamefont {B.}~\bibnamefont {Eckhardt}},\
  }\bibfield  {title} {\enquote {\bibinfo {title} {{Edge of Chaos} in a
  parallel shear flow},}\ }\href {\doibase 10.1103/PhysRevLett.96.174101}
  {\bibfield  {journal} {\bibinfo  {journal} {Phys. Rev. Lett.}\ }\textbf
  {\bibinfo {volume} {96}},\ \bibinfo {pages} {174101} (\bibinfo {year}
  {2006})}\BibitemShut {NoStop}%
\bibitem [{\citenamefont {Cvitanovi{\'c}}\ \emph {et~al.}(2009)\citenamefont
  {Cvitanovi{\'c}}, \citenamefont {Davidchack},\ and\ \citenamefont
  {Siminos}}]{SCD07}%
  \BibitemOpen
  \bibfield  {author} {\bibinfo {author} {\bibfnamefont {P.}~\bibnamefont
  {Cvitanovi{\'c}}}, \bibinfo {author} {\bibfnamefont {R.~L.}\ \bibnamefont
  {Davidchack}}, \ and\ \bibinfo {author} {\bibfnamefont {E.}~\bibnamefont
  {Siminos}},\ }\bibfield  {title} {\enquote {\bibinfo {title} {On the state
  space geometry of the {Kuramoto-Sivashinsky} flow in a periodic domain},}\
  }\href@noop {} {\bibfield  {journal} {\bibinfo  {journal} {SIAM J. Appl. Dyn.
  Syst.}\ }\textbf {\bibinfo {volume} {9}},\ \bibinfo {pages} {1--33} (\bibinfo
  {year} {2009})},\ \bibinfo {note} {\arXiv{0709.2944}}\BibitemShut {NoStop}%
\bibitem [{\citenamefont {Suri}\ \emph {et~al.}(2017)\citenamefont {Suri},
  \citenamefont {Tithof}, \citenamefont {Grigoriev},\ and\ \citenamefont
  {{Schatz}}}]{STGS16}%
  \BibitemOpen
  \bibfield  {author} {\bibinfo {author} {\bibfnamefont {B.}~\bibnamefont
  {Suri}}, \bibinfo {author} {\bibfnamefont {J.}~\bibnamefont {Tithof}},
  \bibinfo {author} {\bibfnamefont {R.~O.}\ \bibnamefont {Grigoriev}}, \ and\
  \bibinfo {author} {\bibfnamefont {M.~F.}\ \bibnamefont {{Schatz}}},\
  }\bibfield  {title} {\enquote {\bibinfo {title} {Forecasting fluid flows
  using the geometry of turbulence},}\ }\href {\doibase
  10.1103/PhysRevLett.118.114501} {\bibfield  {journal} {\bibinfo  {journal}
  {Phys. Rev. Lett.}\ }\textbf {\bibinfo {volume} {118}},\ \bibinfo {pages}
  {114501} (\bibinfo {year} {2017})},\ \bibinfo {note}
  {\arXiv{1611.02226}}\BibitemShut {NoStop}%
\end{thebibliography}%

\end{document}